\documentclass[journal, twocolumn]{IEEEtran}
\usepackage{tikz,tikz-qtree}	
\usepackage{graphicx,cite}
\usepackage{amsmath,amssymb,bbm}
\usepackage{algorithm,algpseudocode}
\usepackage{url,multirow}

\newcommand{\bld}[1]{{\bf #1}}
\newcommand{\bs}[1]{\boldsymbol{#1}}
\newcommand{\inCurly}[1]{\left\lbrace #1 \right\rbrace}
\newcommand{\Expectation}[1]{\mathbb{E}\inCurly{#1} }
\newcommand{\Probability}[1]{ \mathbb{P}\left(#1\right) }
\newcommand{\inBrackets}[1]{\left( #1 \right)}
\newcommand{\inSqBrackets}[1]{\left[ #1 \right]}
\newcommand{\given}{\left|\right.}

\title{Joint spatial filter and time-varying MCLP for dereverberation and interference suppression
	of a dynamic/static speech source}
\author{Srikanth~Raj~Chetupalli,
        and Thippur V. Sreenivas,\\
        Dept. of ECE, Indian Institute of Science, Bangalore, 560012.
        }
\begin{document}

\maketitle

\begin{abstract}
 Dereverberation of a moving speech source in the presence of other directional interferers, is a harder problem than that of stationary source and interference cancellation. We explore joint multi channel linear prediction (MCLP) and relative transfer function (RTF) formulation in a stochastic framework and maximum likelihood estimation. We found that the combination of spatial filtering with distortion-less response constraint, and time-varying complex Gaussian model for the desired source signal at a reference microphone does provide better signal estimation. For a stationary source, we consider batch estimation, and obtain an iterative solution. Extending to a moving source, we formulate a linear time-varying dynamic system model for the MCLP coefficients and RTF based online adaptive spatial filter. For the case of tracking a desired source in the presence of interfering sources, the same formulation is used by specifying the RTF. Simulated experimental results show that the proposed scheme provides better spatial selectivity and dereverberation than the traditional methods, for both stationary and dynamic sources even in the presence of interfering sources.

\end{abstract}
\begin{IEEEkeywords}
Dereverberation, Multi channel linear prediction, Spatial filtering, Linear dynamical system, Relative transfer function.
\end{IEEEkeywords}
%
\section{Introduction}
\label{sec:intro}
For tele-communication, the natural speech inside an enclosure is adversely affected by reverberation due to multiple reflections of the source signal by the walls and other rigid surfaces in the enclosure. This affects hands-free telephony and other man-machine interaction tasks such as HuBot (Human-Robot) interaction, assisted hearing devices and voice assistants relying on distant speech recognition \cite{kuttruff2016room, assmann2004perception,bradley2003importance, gardner1960study, petrick2007harming, lippmann1997speech, nabelek1974monaural}. Enhancement of reverberant speech is a challenging problem due to the long impulse response of enclosure ($0.5-1~s$ typically) and the time varying nature of speech signal properties. The problem is further compounded in the presence of ambient acoustic or recording noise, or interfering sources, and also a dynamic scenario where the source may be moving changing its position over time. In this paper, we consider enhancement of reverberant speech of a static or dynamic source, and also the desired source signal selection in the presence of other directional interferers.
\par In spatial filtering methods such as beamforming, the acoustic propagation between the source and the microphone positions is described using an acoustic transfer function (ATF) or ratio transfer function (RTF) \cite{gannot2001signal} and the late reverb component is treated as noise \cite{capon1969high,habets2010new}. The spatial selective filtering is essential for the suppression of directional interferers and the estimation of source from a desired spatial direction. However, since the diffuse (late reverb) component in the direction of desired source is not suppressed by the spatial filter, it limits dereverberation and interference suppression performance. Blind inverse filtering through delayed multi-channel linear prediction (MCLP) \cite{nakatani2010speech, kinoshita2007linear, integrated2009yoshioka} in the short-time Fourier transform (STFT) domain is an effective approach to reduce reverberation. But MCLP has no spatial selectivity, and the presence of noise and directional interferers causes degradation of prediction filter estimation due to the single source assumption. Hence, a combination of MCLP and spatial filtering with RTF (RTF-MCLP) is considered in this paper, which would be better for dereverberation as well as interference suppression. The RTF can also be estimated, or specified to correspond to a particular direction, to spatially select a desired source in the enclosure.
\par In MCLP, the prediction filters are estimated using a maximum likelihood (ML) criterion with a time-varying power spectral density (PSD) assumption for the prediction residual \cite{nakatani2010speech, integrated2009yoshioka, jukic2015multi, multi2015jukic, late2017chetupalli}. The spatial filters are often designed with a distortion-less response constraint for the desired source direction \cite{cox1987robust, frost1972algorithm, habets2010new}. In the stationary source case, we consider ML estimation with the distortion-less response constraint along with time-domain short-time auto-regressive (AR) model PSD constraint for the desired source. We formulate an iterative solution involving MCLP reverberation cancellation followed by minimum variance distortion-less response (MVDR) spatial filtering, and time-domain AR model based PSD estimation of the spatial filter output (desired signal). In the single source case (no interference), the source RTF is estimated within the iterations of the RTF-MCLP algorithm. Whereas in multi-source (interferer) case, we consider the desired source RTF as known a-priori in the present formulation. It is also possible to estimate the individual source RTFs using constraints such as the direction of arrival, or by restricting RTF estimation to desired source active STFT regions only. 
\par We find that MCLP and spatial filtering can aid each other through the iterative estimation, where MCLP can provide reduced late-reverb noise in the estimated early component signal, the spatial filter can avoid interfering sources from other directions. We extend the joint RTF-MCLP formulation to dynamic sources through the linear dynamical system model for MCLP filters, along with online MVDR spatial filtering. The MCLP filters are estimated using ML criterion in each time frame, and the time-dependent spatial filter for the source RTF is estimated using the output of MCLP. In the single source case, the source RTF is also estimated in each time frame using the MCLP residual.
\subsection{Related works}
\par A single stationary source in a noiseless reverberant environment is considered in the early MCLP formulation of \cite{kinoshita2007linear,nakatani2010speech}, using a time-varying complex Gaussian model and maximum likelihood estimation of the source PSD. Several further works focused on improving the prediction filter estimation using better desired signal models \cite{integrated2009yoshioka, jukic2015multi, multi2015jukic, late2017chetupalli}, and prediction filter models \cite{reverberation2014nicholas,late2017chetupalli}. An extension of the method to noisy measurement case is considered in \cite{integrated2009yoshioka}, and a multi-source case using the blind source separation approach of independent component analysis in \cite{yoshioka2011blind, Otsuka2016Multichannel}. Cascaded spatial filtering is used as a post filter to reduce the residual reverberation in the MCLP enhanced signal in \cite{cohen2017combined, delcroix2014linear}. Though different formulations exist to design a spatial filter \cite{van1988beamforming, capon1969high, frost1972algorithm, Warsitz2007Blind,  habets2010new}, among them MVDR beamformer \cite{capon1969high, habets2010new} is a simple and popular approach.
Spatial filtering provides spatial selectivity, and allows for desired source estimation in recordings with multiple sources. Yet the MCLP dereverberation gets adversely affected in a multi-source case, which results in excess residual reverberation, and hence the post filtering is less effective. Similar to the present work, spatial filtering with in the MCLP iterations using a neural-network (NN) based speech mask \cite{heymann2016neural} is considered in \cite{drude2018integrating}, as a pre-processing step, for an ASR task in noisy recordings. However, no justification is available for the NN mask based approach. Instead, we develop the scheme as a maximum likelihood solution to the joint RTF-MCLP model of reverberant multi microphone signals. 
\par The MCLP speech dereverberation for a dynamic source scenario has been considered in the literature \cite{adaptive2009yoshioka, Togami2013Optimized, adaptive2017jukic, braun2016online}. A recursive least squares (RLS) formulation is proposed in \cite{adaptive2009yoshioka} with source position change detection. Further, it has been extended using a Kalman filter approach and shown to be more effective \cite{braun2016online}. Adaptive estimation of time-varying channel condition is considered in \cite{Togami2013Optimized, adaptive2017jukic}. However, MCLP dereverberation for {\it dynamic sources in the presence of interferers is not considered in the literature}. 
We consider RTF-MCLP along with the Kalman filter based linear dynamical system model for both interference reduction and desired source tracking. We show that the RTF-MCLP is readily extended to dynamic sources, and directional interferences, as online adaptive spatial filtering \cite{higuchi2017online} of signal enhanced through the time-varying MCLP.
\section{RTF-MCLP Formulation}
\label{sec:ProblemFormulation}
Consider an acoustic recording scenario using an $M$ element microphone array and a single sound source inside a reverberant enclosure with no other noise. The signal recorded at the $m^{th}$ microphone $x_m[t]$ can be related to the source signal $s[t]$ using the room impulse response (RIR) relation,
\begin{equation}
	x_m[t] = h_m[t] \circledast s[t].
\end{equation}
$h_m[t]$ being the RIR between source and the $m^{th}$ microphone position and $t$ is the discrete time index. The convolution relationship can be expressed approximately in the STFT domain (ignoring the cross band effects) \cite{avargel2008system} as
\begin{align}\label{eqn:convInSTFT}
	x_m[n,k]&\approx \sum\limits_{p=0}^{P-1} h_m[p,k] s[n-p,k],\\\nonumber
			&=\sum\limits_{p=0}^{D-1} h_m[p,k] s[n-p,k] + \sum\limits_{p=D}^{P-1} h_m[p,k] s[n-p,k],
\end{align}
where $n$ denotes the discrete time index, $0 \leq k \leq K-1$ is the discrete frequency bin index, and $P$ is the length of the RIR in STFT domain, which is equal to the ratio of the duration of time domain impulse response and the window shift used for STFT analysis. Blind estimation of the STFT domain reverb filters $\{h_m[n,k],~\forall~m,k\}$ is difficult, due to the continuous spectro-temporal variation of the speech signal. However, we can split the RIR into two parts: the first part in \eqref{eqn:convInSTFT} modeling the direct component and early reflections, where as the second part models the diffuse late reverb component. We can parameterize the late reverb component using a delayed multi channel linear prediction (MCLP) model\cite{kinoshita2007linear} as below:
\begin{equation}\label{eqn:sigmodel1}
	x_m[n,k] =\underbrace{\sum\limits_{m'=1}^{M} \sum\limits_{l=1}^{L} g_{m,m'}^*[l,k] x_{m'}[n-D-l,k]}_{r_m[n,k]}+d_m[n,k],
\end{equation}
where the delay $D\geq 1$ is chosen to avoid over fitting to speech correlations between successive STFT frames; $D$ controls the duration of the early reflection component retained in prediction residual $d_m[n,k]$. 
\par The late reverb component does vary due to microphone position and hence multi-microphone prediction helps in predicting out the diffuse component. The predicted component $r_m[n,k]$ in \eqref{eqn:sigmodel1} at the $m^{th}$ microphone can be expressed compactly in vector form as,
\begin{equation}\label{eqn:mclpVecForm}
	r_{m}[n,k] = \bld{g}_{m}[k]^H \bs{\phi}[n,k].
\end{equation}
The vector of predictor STFT samples is:
\begin{multline}
	\bs{\phi}[n,k] = \left[ x_1[n-D-1,k],\dots,x_1[n-D-L,k],
	\right.\\\left. \dots, x_M[n-D-1,k],\dots,x_M[n-D-L,k] \right]^T,
\end{multline}
and the vector of MCLP coefficients is given by
\begin{multline}
\bld{g}_{m}[k]= \left[ g_{m,1}[0,k],\dots,g_{m,1}[L-1,k],
	\right.\\\left. \dots, g_{m,M}[0,k],\dots,g_{m,M}[L-1,k] \right]^T.
\end{multline}
Let $\bld{r}[n,k]=\left[r_1[n,k],\dots,r_M[n,k]\right]^T$ be the vector form for the late reverb component at all the $M$ microphones, which can be written as,
\begin{gather}\label{eqn:lateComponentModel}
	\bld{r}[n,k]=\bld{G}[k]^H \bs{\phi}[n,k],~\bld{G}[k]=\left[\bld{g}_{1}[k]~\dots \bld{g}_{M}[k]\right].
\end{gather}
The MCLP parameters $\{\bld{G}[k]\}$ can be estimated jointly using a stochastic formulation.
\par Now, to include the directionality of desired source, we approximate the desired early component (first component in eqn. \eqref{eqn:convInSTFT}) using a multiplicative transfer function (for a small $D$) as
\begin{equation}
	d_{m}[n,k] \approxeq H_{m}[k] s[n,k],~\forall~m.
\end{equation}
In vector notation, we can write, 
\begin{equation}\label{eqn:earlyComponentModel}
	\bld{d}[n,k] \approxeq \bld{h}[k] s[n,k]
\end{equation}
where $\bld{h}[k]=[H_{1}[k] \dots H_{M}[k]]^T$ is the stack of each mic-to-source transfer function (only early reflections), and $s[n,k]$ is the scalar signal STFT.
\par Using \eqref{eqn:earlyComponentModel} and \eqref{eqn:lateComponentModel}, we can write the total signal model as,
\begin{equation}\label{eqn:earlyLateModel}
\bld{x}[n,k]=\bld{h}[k] s[n,k]+\bld{G}[k]^H \bs{\phi}[n,k],
\end{equation}
comprising early reflection part, and the predictable late reverb component. Since we need to estimate the source signal $s[n,k]$, and we don't known $\bld{h}[k]$ also, we can resort to a relative transfer function (RTF) approach using a reference microphone. i.e., the first part of \eqref{eqn:earlyLateModel} can be expressed using the RTF with respect to a reference microphone say $r=1$:
\begin{equation}\label{eqn:retfSignalModel}
\bld{x}[n,k]=\bld{a}[k] d_{1}[n,k]+\bld{G}[k]^H \bs{\phi}[n,k],
\end{equation}
where 
\begin{equation}
	\bld{a}[k]= \left[1~\frac{H_{2}[k]}{H_{1}[k]} \dots \frac{H_{M}[k]}{H_{1}[k]} \right]^T.
\end{equation}
Let us choose a weight vector ${\bld{w}[k]}$ (spatial filter) such that ${\bld{w}}[k]^H\bld{a}[k]=1$. We can re-write \eqref{eqn:retfSignalModel} as,
\begin{equation}\label{eqn:beamformedModel}
	{\bld{w}}[k]^H \bld{x}[n,k]=d_{1}[n,k]+{\bld{w}}[k]^H\bld{G}[k]^H \bs{\phi}[n,k].
\end{equation}
Now, we can consider the estimation of signal component $d_{1}[n,k]$ (desired signal) at the reference microphone. Using the multi channel observations $\{x_m[n,k],~0\leq n \leq N-1,~1\leq m\leq M\}$, a batch method is presented in sec. \ref{sec:proposedMethod}, which is further extended to online estimation in sec. \ref{sec:onlineMethod}. For the ML solution, we consider the frequency bins to be statistically independent and the cross-frequency effects be considered negligible. Thus, the frequency bin index $k$ is omitted for brevity in the formulation below.
\section{Static Source Batch Estimation}
\label{sec:proposedMethod}
\subsection{Gaussian model of early reflection component}
Consider $d_1[t]$, which is the inverse STFT of $d_1[n,k]$, to be the speech signal with early reflections. Let $d_1[t]$ be the output of a quasi-stationary auto regressive model of order $Q$, and let $\inCurly{\gamma_{nk},~\forall~k}$ denote the time-varying power spectral density (PSD) of $d_1[t]$ at the STFT frame index $n$ and frequency index $k$. We consider the signal $\{d_1[n,k],\forall~k\}$ to be independent (across the time and frequency indices $n,k$), circularly symmetric complex Gaussian random variable with variance $\gamma_{nk}$; i.e., $\Probability{\bld{d}_1[k]} = \prod\limits_{n=1}^{N} \Probability{d_1[n,k]}$, and
\begin{equation}\label{eqn:ecModel}
	\Probability{d_1[n,k]}= \mathcal{N}_c\inBrackets{d_1[n,k];{0}, \gamma_{nk}}.
\end{equation}
The independence ignores time dependence of speech PSD, but the frequency dependence is imposed using time domain AR model. The discussion below is similar for each frequency bin $k$, and hence we have omitted the index $k$ for brevity.
\par Using \eqref{eqn:beamformedModel} and \eqref{eqn:ecModel}, we can express the stochastic model of $\bld{x}[n]$ as:
\begin{multline}
	\Probability{\bld{x}[n] \given {\bld{w}},\bld{G},\bs{\phi}[n],{\gamma}_{n}} =\\ \frac{1}{\pi {\gamma}_{n}}  \exp\inBrackets{-{\gamma}_n^{-1} \left| {\bld{w}}^H \left(\bld{x}[n]-\bld{G}^H \bs{\phi}[n] \right) \right|^2  },
\end{multline}
and
\begin{equation}
	\Probability{\bld{X}\given \bs{\Theta}}=\prod\limits_{n=0}^{N-1} \Probability{\bld{x}[n]\given {\bld{w}},\bld{G},\bs{\phi}[n],{\gamma}_n},
\end{equation}
where $\bs{\Theta}=\{\bld{w},\bld{G},\bs{\gamma}=\{{\gamma}_n,~\forall~n\} \}$ is the set of all parameters to be estimated given the multi-microphone signals $\bld{X} = \inSqBrackets{\bld{x}[0],\dots,\bld{x}[N-1]}$.
\subsection{Parameter Estimation}
Using the above stochastic formulation of $\bld{x}[n]$, we consider maximum likelihood (ML) parameter estimation:
\begin{equation}
	\hat{\bs{\Theta}}=\underset{\bs{\Theta}}{\arg\max}~~\Probability{ \bld{X} \given \bs{\Theta} }.
\end{equation}
Equivalently, negative logarithm of the likelihood function can be minimized. i.e., $\mathcal{L}(\bs{\Theta}) \propto -\log p(\bld{X}|\bs{\Theta})$, where
\begin{gather}\label{eqn:llk}
	\hspace{-4pt}
 \mathcal{L}(\bs{\Theta}) \triangleq \sum\limits_{n=0}^{N-1} \left[ \log\gamma_n +
 	\gamma_n^{-1} \left| {\bld{w}}^H \left(\bld{x}[n]-\bld{G}^H \bs{\phi}[n] \right) \right|^2 \right].
\end{gather}
Direct minimization of $\mathcal{L}(\bs{\Theta})$ with respect to the variables $\{\gamma_n,~\forall~n\}$, $\bld{w}$ and $\bld{G}$ is not possible, because of the inter-dependencies between them. Instead, we consider a coordinate descent approach, in which the variables are successively optimized in a sequential, iterative manner. 
\subsubsection{Estimation of MCLP filters ($\bld{G}$)}
\par The part of cost function $\mathcal{L}(\bs{\Theta})$ relevant for the estimation of $\bld{G}$ is,
\begin{equation}\label{eqn:llkG}
\mathcal{L}(\bld{G}) =  \sum\limits_{n=0}^{N-1}	\hat{\gamma}_n^{-1} \left| \hat{\bld{w}}^H \left(\bld{x}[n]-\bld{G}^H \bs{\phi}[n] \right) \right|^2.
\end{equation}
Since we are estimating the parameters sequentially, let $\hat{\gamma}_n$ and $\hat{\bld{w}}$ be the estimates from previous iteration. Using the following definitions
\begin{eqnarray}
	\bld{R}_{\phi\phi}\triangleq \sum\limits_{n=0}^{N-1} \hat{\gamma}_n^{-1} \bs{\phi}[n]\bs{\phi}[n]^H,
	~\bld{R}_{\phi x}\triangleq \sum\limits_{n=0}^{N-1} \hat{\gamma}_n^{-1} \bs{\phi}[n] \bld{x}[n]^H,
\end{eqnarray}
and $\bld{W}\triangleq {\bld{w}}{\bld{w}}^H$, the term dependent on MCLP filter $\bld{G}$ in the likelihood function of eqn. \eqref{eqn:llkG} can be written, using the trace notation, as:
\begin{equation}\label{eqn:mlCostG}
\mathcal{L(\bld{G})} = tr\left( \left[ \bld{G}^H  \bld{R}_{\phi \phi} \bld{G}  - \bld{R}_{\phi x}^H \bld{G} - \bld{G}^H  \bld{R}_{\phi x } \right] \bld{W} \right).
\end{equation}
Taking the derivative of eqn. \eqref{eqn:mlCostG} with respect to the matrix variable $\bld{G}$ and equating to zero, we get
\begin{equation}
\bld{W}	\left[\bld{G}^H \bld{R}_{\phi \phi}- \bld{R}_{\phi x}^H \right]=\bld{0}.
\end{equation}
The solution for this is not unique, since $\bld{W}=\bld{w}\bld{w}^H$ is a rank-one matrix, and any matrix in the null space of $\bld{W}$ will satisfy the above equality. However, since it is iterative optimization, we can consider an approximate solution of
\begin{equation}
\left[\bld{G}^H \bld{R}_{\phi \phi}- \bld{R}_{\phi x}^H \right]=\bld{0},
\end{equation}
to lead to a useful solution through the iterations. The MCLP filter matrix $\bld{G}$ solution is obtained as, 
\begin{equation}\label{eqn:gest}
	\hat{\bld{G}}=  \bld{R}_{\phi\phi}^{-1} \bld{R}_{\phi x}.
\end{equation}
i.e., the estimate for $\hat{\bld{G}}$ is obtained similar to the WPE method \cite{nakatani2010speech}, but as we show later in sec. \ref{sec:gammaEstimation}, solution for the weight parameter ${\gamma}_n$ is different in the present approach.
\subsubsection{Steering Vector $(\bld{a})$}
\par Consider the prediction residual computed using the estimate for $\bld{G}$,
\begin{equation}\label{eqn:predresest}
	\hat{\bld{d}}[n] = \bld{x}[n] - \hat{\bld{G}}^H \bs{\phi}[n].
\end{equation}
We note that, the estimate $\hat{\bld{d}}[n]$ comprises of the desired early component signal $\bld{d}[n] = \bld{a} d_1[n]$ and also the residual reverberation $\tilde{\bld{r}}[n]$. Assuming the two components to be uncorrelated, we can write,
\begin{align}\label{eqn:predrescor}
	\bld{R}_{\bld{d}\bld{d}} & = \mathbb{E} \inCurly{\hat{\bld{d}}[n]\hat{\bld{d}}[n]^H}\nonumber\\
	& \approx \bld{a}\bld{a}^H \mathbb{E} \inCurly{|d_1[n]|^2}  + \mathbb{E} \inCurly{\tilde{\bld{r}}[n]\tilde{\bld{r}}[n]^H}.
\end{align}
Because the energy of residual reverberation $\tilde{\bld{r}}[n]$ is smaller than that of desired signal, we can ignore it and we can estimate the RTF $\bld{a}$ from first column of the correlation matrix \cite{gannot2001signal}:
\begin{equation}\label{eqn:hest}
	\hat{\bld{a}} = \frac{ \bld{R}_{\bld{d}\bld{d}} \bld{e}_1 }{\bld{e}_1^H \bld{R}_{\bld{d}\bld{d}} \bld{e}_1},
\end{equation}
where $\bld{e}_1$ is the first column of the $M \times M$ identity matrix. A different estimate based on the eigen vector corresponding to largest eigen value of $\bld{R}_{\bld{d}\bld{d}}$ can also be used \cite{markovich2009multi}. (Different methods of computing RTF are not considered at present).
\subsubsection{Estimation of Spatial Filter $(\bld{w})$}
\par The term dependent on spatial filter $\bld{w}$ in the likelihood function \eqref{eqn:llk} is,
\begin{equation}
	\mathcal{L}(\bld{w})= \bld{w}^H \left[ \sum\limits_{n=1}^{N} \hat{\gamma}_n^{-1} \hat{\bld{d}}[n] \hat{\bld{d}}[n]^H \right] \bld{w} \triangleq \bld{w}^H \bld{R}_{\hat{\bld{d}} \hat{\bld{d}} } \bld{w}.
\end{equation}
Using the distortion-less response constraint of \eqref{eqn:beamformedModel}, the ML estimation problem can be stated as,
\begin{gather}\label{eqn:mvdrcriterion}
	\mbox{minimize}~\inCurly{ \bld{w}^{H} \bld{R}_{\hat{\bld{d}} \hat{\bld{d}} } \bld{w} },\mbox{ s.t. } \bld{w}^H\hat{\bld{a}}=1.
\end{gather}
This is similar to the MVDR formulation, where $\bld{R}_{\hat{\bld{d}} \hat{\bld{d}} }$ denotes the spatial covariance matrix of the filter input. Solution to \eqref{eqn:mvdrcriterion} can be obtained as \cite{habets2010new},
\begin{equation}\label{eqn:west}
	\hat{\bld{w}}=\frac{ \bld{R}_{\hat{\bld{d}} \hat{\bld{d}} }^{-1} \hat{\bld{a}} }{\hat{\bld{a}}^H \bld{R}_{\hat{\bld{d}} \hat{\bld{d}} }^{-1} \hat{\bld{a}} }.
\end{equation}
As noted earlier, the estimated early reflection signal component $\hat{\bld{d}}[n]$ also includes some amount of residual reverberation $\tilde{\bld{r}}[n]$ (since it is diffuse). Using the model in \eqref{eqn:retfSignalModel}, the optimization criterion \eqref{eqn:mvdrcriterion} can be re-stated as,
\begin{gather}\label{eqn:mvdrcriterion2}
	\mbox{minimize}~\inCurly{ \bld{w}^{H} \bld{R}_{\tilde{\bld{r}} \tilde{\bld{r}} } \bld{w} },\mbox{ s.t. } \bld{w}^H\hat{\bld{a}}=1.
\end{gather}
We assume the spatial correlation matrix of the residual reverberation component $\bld{R}_{\tilde{\bld{r}} \tilde{\bld{r}} }$ to be proportional to the spatial correlation matrix of the predicted reverberation component $\hat{\bld{r}}[n] = \hat{\bld{G}}^H \bs{\phi}[n]$, i.e., 
\begin{equation}
	\bld{R}_{\tilde{\bld{r}} \tilde{\bld{r}} } \propto \bld{R}_{\hat{\bld{r}} \hat{\bld{r}} } = \sum\limits_{n=1}^{N} \hat{\gamma}_n^{-1} \hat{\bld{r}}[n] \hat{\bld{r}}[n]^H.
\end{equation}
The above assumption is justified since the reverberation component and the residual reverberation component are both assumed to be diffuse and spatially homogeneous \cite{cohen2017combined}, in general sound fields. Thus, the corresponding spatial filter solution is,
\begin{equation}\label{eqn:west2}
	\hat{\bld{w}}=\frac{ \bld{R}_{\hat{\bld{r}} \hat{\bld{r}} }^{-1} \hat{\bld{a}} }{\hat{\bld{a}}^H \bld{R}_{\hat{\bld{r}} \hat{\bld{r}} }^{-1} \hat{\bld{a}} }.
\end{equation}
We use the spatial filter in \eqref{eqn:west2} for experimentation in this paper, since the solution in \eqref{eqn:west} is found to be sensitive to errors in RTF estimation $\hat{\bld{a}}$.
\subsubsection{Estimation of Desired Signal Variance $(\gamma_n)$}\label{sec:gammaEstimation}
\par The part of objective function of \eqref{eqn:llk} dependent on the STFT signal variance $\gamma_n$ is,
\begin{equation}
	\mathcal{L}(\gamma_n)=\log \gamma_n +\gamma_n^{-1} |\hat{d}_1[n]|^2,~\forall n,
\end{equation}
where $\hat{d}_1[n] = \hat{\bld{w}}^H \left( \bld{x}[n]-\hat{\bld{G}}^H \bs{\phi}[n] \right)$ is the spatial filtered MCLP residual. Minimizer of the objective function is the energy of signal $\hat{d}_1[n]$, i.e., $|\hat{d}_1[n]|^2$. This is an instantaneous estimate for each STFT bin $k$. We can make the variance estimate more consistent by exploiting the correlation across frequency bins, i.e., estimating a smoothed PSD of $d_1[n]$. We resort to time-domain low-order AR modeling of the residual signal $\hat{d}_1[n,k]$, which is obtained after computing the spatial filtered residue signal for each frequency bin $k$. We choose a fixed predictor order of $Q=21$ for the AR model, and the estimated PSD at time index $n$ is given by \cite{Makhoul1975Linear}
\begin{equation}\label{eqn:armodelest}
	H_n[z]=\frac{G_n}{1-\sum\limits_{q=1}^{Q} a_{pn} z^{-p}}.
\end{equation}
The frequency response of the time domain AR model $H_n[z]$ is used to derive the estimate for $\gamma_n$ as
\begin{equation}\label{eqn:gammaest}
	\hat{\gamma}_{nk}={\left|H_n[z_k]\right|^2},~z_k= e^{-j2\pi k/K}.
\end{equation}
We found that this estimate is better than the direct ML estimate $\left| \hat{d}_1[n,k] \right|^2$ for $\gamma_{nk}$.
\subsection{Iterative Parameter Estimation of RTF-MCLP}\label{sec:algoDescription}
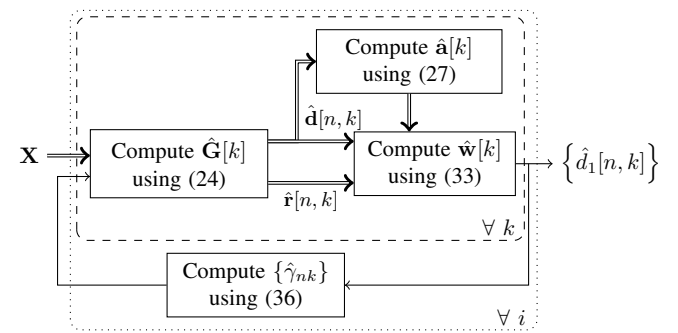
\begin{figure}[h]
	\centering
	\begin{tikzpicture}[scale=0.85, every node/.style={scale=0.85}]
		\node[] (inNode) at (-1.0in,0.05in) {${\bf X}$}; 
		\node[draw,text width=1in,align=center] (filter) at (-0.2,0) {Compute $\hat{{\bf G}}[k]$ using \eqref{eqn:gest}};		
		\node[draw,text width=1.05in,align=center] (filtest) at (3.4,1.6) {Compute $\hat{\bld{a}}[k]$ using \eqref{eqn:hest} };		
		\draw[double,->] ([yshift=3.5mm] filter.east) -| node[right,pos=0.65]{\small $\hat{\bld{d}}[n,k]$} ([xshift=-3mm] filtest.west) --  (filtest.west);		
		\node[draw,text width=0.9in,align=center] (residual) at (3.8,0) {Compute $\hat{\bld{w}}[k]$ using \eqref{eqn:west2}};		
		\draw[double,->] (filtest) -- (filtest |- residual.north);
		\node[draw,text width=1.0in,align=center] (psd) at (1,-1.9) {Compute $\{\hat{\gamma}_{nk}\}$ using \eqref{eqn:gammaest}};

		\draw[double,->] ([yshift=3.5mm]filter.east) --  ([yshift=3.5mm]residual.west);		
		\draw[double,->] ([yshift=-3mm]filter.east) --node[below]{\small $\hat{\bld{r}}[n,k]$}  ([yshift=-3mm]residual.west);
		
		\draw[->] ([xshift=2mm]residual.east) |- (psd.east);
		\draw[->] (psd.west) -- (-2.1,-1.9) |- ([yshift=-2mm]filter.west);
		\draw[double,->] (inNode) -- (inNode -| filter.west);
		\node[] (outNode) at (6.55,0) {$\inCurly{\hat{d}_1[n,k]}$};
		
		\draw[dashed, rounded corners] (-1.8,-1.2) rectangle (5.2,2.3);
		\draw[dotted, rounded corners] (-1.9,-2.6) rectangle (5.4,2.4);
		\node[] at (5.0,-2.4) {$\forall~i$};
 		\node[] at (4.8,-1.0) {$\forall~k$};
		\draw[->] (residual.east) -- (outNode);
	\end{tikzpicture}
	\vspace{-5pt}	
	\caption{Block diagram of the RTF-MCLP algorithm}\label{fig:blockdiagram}
\end{figure}
A block diagram of the joint estimation of MCLP and RTF, iteratively, is shown in Fig. \ref{fig:blockdiagram}. At iteration $i$, we utilize the past estimate of desired speech PSD $\hat{\gamma}_{nk}$ and compute the MCLP filter using \eqref{eqn:gest}. Then the prediction residual $\hat{d}_m[n,k]$ for each microphone $m$ is computed for each frequency bin $k$. This prediction residual is then used to estimate the RTF $\bld{a}[k]$, and the spatial filter $\bld{w}[k]$ and desired signal $\hat{d}_1[n,k]$ are obtained. The output of spatial filter $\hat{d}_1[n,k]$ is then used to compute the desired signal PSD for the next iteration. The reverberant signal $\bld{x}[n,k]$ itself is considered as initialization of $\hat{\bld{d}}[n,k]$ for the first iteration, and the initial estimates for steering vector and the spatial filter are computed. Output of the spatial filter is then used to compute initial estimate of desired signal PSD $\gamma_{nk}$.
\subsection{Desired Source Selection of Multiple Sources}
In \eqref{eqn:retfSignalModel}, we have considered a single source and a spatial filter to enhance the MCLP performance. Let us now consider multiple sources $s=1,2,\dots,S$ contributing to the $M$ microphone signals. However, the diffuse late reverb component due to all the $S$ sources is predictable by MCLP \cite{yoshioka2011blind}, the prediction residual contains early reflection components due to all the sources. Thus,
\begin{equation}
	\bld{x}[n,k] = \sum_{s=1}^{S} \bld{a}_s[k] d_{s1}[n,k] + \bld{r}[n,k],
\end{equation}
where $\bld{a}_s[k]$ is the RTF vector for the source $s$. When we want to estimate a desired source ($s^*$) among the multiple sources, we assume the corresponding RTF to be known a-priori, i.e., $\bld{a}_{s^*}[k]$ is known\footnote{It is possible to select a desired source using other criteria such as dominant energy, etc.}. Using the algorithm described in sec. \ref{sec:algoDescription}, the desired signal is estimated subject to the constraint $\bld{w}[k]^H\bld{a}_{s^*}[k]=1$. Since the PSD $\gamma_{nk}$ is estimated after spatial filtering, it corresponds to the desired source through successive iterations. MCLP estimation with the constraint of desired source signal PSD as the estimate for $\gamma_{nk}$ improves estimation of the early component iteratively. The noise spatial covariance matrix computed for spatial filtering in \eqref{eqn:mvdrcriterion2} includes the late reverb components of desired source and the interferers. Hence, minimization of noise power in spatial filtering suppresses the interfering sources along with the residual reverberation of the desired source. 
\section{Dynamic Source Online Estimation}
\label{sec:onlineMethod}
The batch method described earlier can be considered as comprising of two stages: the MCLP dereverberation and distortion-less response spatial filtering. The two stages aid each other through the iterations, one reducing the effect of diffuse reverb component and the other reducing the effect of interfering source/noise. The advantage of this joint optimization can be exploited for spatial filtering of a moving source also. But, the moving source based MCLP filtering requires a time-varying MCLP model, so that the predictors are optimally adapted. The issue to be resolved for a moving source is that of changing the RTF with respect to microphones and further its effect on the MCLP itself.
\par Since our static analysis is based on STFT, even for the moving source case, the same linear signal model is used with a slight generalization:
\begin{equation}
	x_m[n,k] = d_m[n,k] + \bld{g}_{m,n}[k]^H \bs{\phi}[n,k],
\end{equation}
where the MCLP coefficients $\bld{g}_{m,n}[k]$ are dependent on the STFT frame index $n$. In vector form, we can express the RTF-MCLP model as
\begin{equation}
	\bld{x}[n,k] = \bld{a}_n[k]d_1[n,k] + \bld{G}_{n}[k]^H \bs{\phi}[n,k].
\end{equation}
We note that the direct ML estimation using the distortion-less response constraint for the spatial filter, and a linear dynamical system model for MCLP will be intractable. Hence, we consider the desired signal component $\bld{d}[n,k] = \bld{a}_n[k] d_1[n,k]$ to be an isotropic complex Gaussian random variable, $\Probability{\bld{d}[n,k]} = \mathcal{N}_c \inBrackets{ \bld{0},\gamma_{nk} \bld{I} }$, and hence
\begin{equation}\label{eqn:sigmodeltv}
	\Probability{\bld{x}[n,k]} = \mathcal{N}_c \inBrackets{ \bld{G}_n[k]^H\bs{\phi}[n,k] ,\gamma_{nk} \bld{I} }.
\end{equation}
Now we formulate an optimum time-varying MCLP using a linear dynamic system approach and also perform online spatial filtering; we can then estimate the variable $\gamma_{nk}$ using the spatial filter output similar to the batch method.
\subsection{Time-varying MCLP}\label{sec:ldsModel}
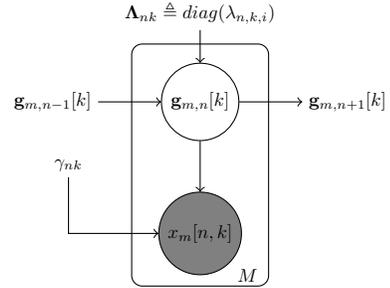
\begin{figure}[h]
	\centering
	\begin{tikzpicture}[scale=0.7,every node/.style={scale=0.7}]

		\draw[rounded corners] (1.4,-2) rectangle (4,2.6);

		\node[] (gmn) at (-0.1,1.5) {$\bld{g}_{m,n-1}[k]$};
		
		\node[] (lmn2) at (2.7,3.2) {$\bs{\Lambda}_{nk}\triangleq diag(\lambda_{n,k,i}) $};
		\node[draw,circle,fill=gray,minimum width=12mm] (xmn2) at (2.7,-1) {$x_m[n,k]$};
		\node[draw,circle,minimum width=12mm] (gmn2) at (2.7,1.5) {$\bld{g}_{m,n}[k]$};
		\draw[->] (lmn2) -- (gmn2);
		\draw[->] (gmn2) -- (xmn2);
		\draw[->] (gmn) -- (gmn2);
		\node[] (p2) at (0.2,0.3) {$\gamma_{nk}$};
		\draw[->] (p2) |- (xmn2.west);
		
		\node[] (gmn3) at (5.5,1.5) {$\bld{g}_{m,n+1}[k]$};
		\draw[->] (gmn2) -- (gmn3);
		\node[] at (3.6,-1.8) {$M$};
	\end{tikzpicture}
	\caption{Graphical model of the $M$ microphone observations.}
\end{figure}
We consider a Gaussian innovation model for the time-varying MCLP coefficients $\{\bld{g}_{m,n}[k],~\forall~m\}$ given by
\begin{equation}\label{eqn:coefficientModel}
	\bld{g}_{m,n}[k]= \bld{g}_{m,n-1}[k]+\bld{i}_{m,n}[k],~\forall~m\in \inSqBrackets{1~M},
\end{equation}
where the innovation term $\bld{i}_{m,n}[k]$ is a complex Gaussian distributed random variable with mean zero and covariance matrix $\bs{\Lambda}_{nk}$. The innovation is i.i.d (independent and identically distributed) across all the microphones, i.e., $\bld{i}_{m,n}[k]=\mathcal{N}_c (\bld{i}_{m,n}[k]; \bld{0},\bs{\Lambda}_{nk}),~\forall~m$. We consider a diagonal covariance matrix for the innovation term, for the sake of simplicity $\bs{\Lambda}_{nk}\triangleq diag(\lambda_{n,k,i})$. Thus, the prediction coefficients can vary independently for  each frequency bin, each microphone $m$ and time frame $n$. As earlier, the frequency bin index $k$ is omitted below for brevity.
\par We consider maximum likelihood criterion for estimating the time-varying parameters. At time frame $n$, we consider the optimization problem,
\begin{equation}\label{eqn:mlcostdyn}
	\mbox{maximize}~
	 \inCurly{ \log \Probability{ \bld{X}_{0:n} \given \bs{\Theta}_{n} } },
\end{equation}
where $\bld{X}_{0:n} = \inSqBrackets{\bld{x}[0],\dots,\bld{x}[n]}$, $\bs{\Theta}_{n} =\bs{\theta}_{0:n}$, where $\bs{\theta}_n=\{\bs{\Lambda}_{n},{\gamma}_{n}\}  $. We consider a time recursive solution, 
\begin{equation}\label{eqn:mlcostdyn2}
\mbox{maximize}~\inCurly{ \log \Probability{ \bld{x}[n] \given \bld{X}_{0:n-1}, \hat{\bs{\Theta}}_{n-1}, \bs{\theta}_n } }.
\end{equation}
The problem can be solved sequentially using the EM algorithm \cite{dempster1977maximum}. The E-step requires computation of the posterior distribution $\Probability{\bld{G}_{n} \given \bld{X}_{0:n}, \bs{\Theta}_{n} }$.
Using the i.i.d assumption in \eqref{eqn:sigmodeltv}, \eqref{eqn:coefficientModel} over the microphone index $m$, we can show that
\begin{equation}
\Probability{\bld{G}_{n} \given \bld{X}_{0:n}, \bs{\Theta}_{n} } = \prod_{m=1}^{M} \Probability{\bld{g}_{m,n} \given \bld{X}_{0:n}, \bs{\Theta}_{n} },\mbox{where}\nonumber
\end{equation}
\begin{equation}\label{eqn:dynJoint2}
 \Probability{\bld{g}_{m,n} \given \bld{X}_{0:n}, \bs{\Theta}_{n} } \propto \Probability{x_m[n],\bld{g}_{m,n} \given\bld{X}_{0:n-1}, \bs{\Theta}_{n} },
\end{equation}
and
\begin{multline}
	\Probability{x_m[n],\bld{g}_{m,n} \given\bld{X}_{0:n-1}, \bs{\Theta}_{n} } = \\ \Probability{x_m[n] \given \bld{g}_{m,n}, \gamma_n } \Probability{\bld{g}_{m,n} \given \bld{X}_{0:n-1}, \bs{\Theta}_{n} }.
\end{multline}
The distribution $\Probability{\bld{g}_{m,n} \given \bld{X}_{0:n-1},\bs{\Theta}_{n}}$ is obtained after marginalizing over the random variable $\bld{g}_{m,n-1}$, and it corresponds to a Gaussian distribution. Let
\begin{equation}
	\Probability{\bld{g}_{m,n} \given \bld{X}_{0:n-1},\bs{\Theta}_{n}} = \mathcal{N}_c \left( \bs{\mu}_{m,n|n-1}, \bs{\Sigma}_{n|n-1} \right),
\end{equation}
using \eqref{eqn:coefficientModel}, we can show that
\begin{eqnarray}\label{eqn:kalmanPrior}
	\bs{\mu}_{m,n\given n-1}=\bs{\mu}_{m,n-1\given n-1},~
	\bs{\Sigma}_{n\given n-1}= \bs{\Sigma}_{n-1\given n-1} +\bs{\Lambda}_n,
\end{eqnarray}
i.e., mean is unchanged but the covariance is getting corrected. 
Similarly, we have the posterior distribution 
\begin{gather}\label{eqn:post1}
	\Probability{\bld{g}_{m,n} \given \bld{X}_{0:n},\bs{\Theta}_{n}}= \mathcal{N}_c\inBrackets{ \bs{\mu}_{m,n|n}, \bs{\Sigma}_{n|n}},\\
	\mbox{where}~\bs{\Sigma}_{n \given n}=\left[ \bs{\Sigma}_{n\given n-1}^{-1} + \gamma_n^{-1} \bs{\phi}[n]\bs{\phi}[n]^H \right]^{-1},~\mbox{and}\nonumber\\\nonumber
	\bs{\mu}_{m,n\given n}=\bs{\Sigma}_{n\given n} \left[ \bs{\Sigma}_{n\given n-1}^{-1} \bs{\mu}_{m,n\given n-1} + \gamma_n^{-1} x_m^*[n] \bs{\phi}[n]  \right],
\end{gather}
where $x_m^*[n]$ is the complex conjugate of the scalar $x_m[n]$. The equations in \eqref{eqn:post1} can be simplified and rewritten \cite{bishop2006pattern} as,
\begin{eqnarray}\label{eqn:kalmanPost}
	{e}_m[n]\triangleq x_m[n]-\bs{\mu}_{m,n|n-1}^H \bs{\phi}[n],~\forall~m\\
	\bld{K}[n] = \frac{ \bs{\Sigma}_{n|n-1} \bs{\phi}[n]  }{\gamma_n + \bs{\phi}[n]^H \bs{\Sigma}_{n|n-1} \bs{\phi}[n] },\\
	\bs{\Sigma}_{n|n}=\left[ \bld{I}-\bld{K}[n] \bs{\phi}[n]^H \right]\bs{\Sigma}_{n|n-1},\\
	\bs{\mu}_{m,n|n}=\bs{\mu}_{m,n|n-1} + \bld{K}[n] {e}_m^*[n],~\forall~m.
\end{eqnarray}
The computational expressions are similar to the traditional Kalman filter estimation, for each $m$. The posterior mean is taken as the estimated prediction filter at time frame $n$,
\begin{equation}\label{eqn:filtest}
	\hat{\bld{g}}_{m,n}\triangleq \bs{\mu}_{m,n|n},~\forall~m,
\end{equation}
and the estimates of early and the late reverb components are computed as:
\begin{eqnarray}\label{eqn:sigest}
\hat{{r}}_m[n] = \hat{\bld{g}}_{m,n}^H \bs{\phi}[n],~\mbox{and }
\hat{{d}}_m[n]={x}_m[n]-\hat{{r}}_m[n].
\end{eqnarray}
The computations in \eqref{eqn:kalmanPrior}, \eqref{eqn:kalmanPost} require estimates of the parameters $\gamma_n,\bs{\Lambda}_n$. We use the estimate $\hat{\bs{\Lambda}}_{n-1}$ of the previous time step in the computations of $\eqref{eqn:kalmanPrior}$, and the parameter $\gamma_n$ is computed after online spatial filtering similar to eqn. \eqref{eqn:gammaest}.
\subsection{Online spatial filtering}
The optimization problem in eqn. \eqref{eqn:mvdrcriterion2} can be modified for the online estimation case as,
\begin{align}\label{eqn:mvdrCost}
	\hat{\bld{w}}_n = \underset{\bld{w}_n}{\arg\min}~
	\bld{w}_n^H \bld{R}_{\hat{\bld{r}} \hat{\bld{r}} }[n] \bld{w}_n,
\end{align}
subject to the constraint $\bld{w}_n^H \bld{a}_n = 1$. A running estimate of the covariance matrix $\bld{R}_{\hat{\bld{r}} \hat{\bld{r}} }[n]$ is computed using recursive averaging with smoothing constant $\alpha_1$ as,
\begin{equation}\label{eqn:noisecovarupdate}
\bld{R}_{\hat{\bld{r}} \hat{\bld{r}} }[n] = (1-\alpha_1) \bld{R}_{\hat{\bld{r}} \hat{\bld{r}} }[n-1] + \alpha_1 \gamma_n^{-1} \hat{\bld{r}}[n] \hat{\bld{r}}[n]^H.
\end{equation}
The solution to \eqref{eqn:mvdrCost} is computed similar to \eqref{eqn:west2},
\begin{equation}\label{eqn:mvdrSolution}
	\hat{\bld{w}}_n = \frac{\bld{R}_{\hat{\bld{r}} \hat{\bld{r}} }[n]^{-1} \bld{a}_n}{\bld{a}_n^H \bld{R}_{\hat{\bld{r}} \hat{\bld{r}} }[n]^{-1} \bld{a}_n}.
\end{equation}
\subsection{Parameter Estimation}
\subsubsection{Gaussian Innovation Covariance}
\par The matrix $\bs{\Lambda}_n = diag( {{\lambda}}_{n,i} )$ is the covariance of innovations in MCLP coefficients, which can be estimated as
\begin{equation}
	{{\lambda}}_{n,i} = \frac{1}{M} \sum\limits_{m=1}^{M} \Expectation{ \left| \bld{g}_{m,n}[i] - 	\bld{g}_{m,n-1}[i] \right|^2 },
\end{equation}
where $\bld{g}_{m,n}[i]$ is the $i^{th}$ entry of the vector $\bld{g}_{m,n}$, and the expectation is taken with respect to the joint distribution $\Probability{ \bld{G}_n,\bld{G}_{n-1} \given \bld{X}_{0:n},\bs{\Theta}_n }$. To avoid the computation of joint distribution, which requires backward recursion \cite{bishop2006pattern}, we consider a simpler estimate based on the changing mean vector:
\begin{equation}
	\hat{{\lambda}}_{n,i} = \frac{1}{M} \sum_{m=1}^{M} \left| \bs{\mu}_{m,n|n}[i] - \bs{\mu}_{m,n-1|n-1}[i] \right|^2 + \epsilon,
\end{equation}
where $\epsilon$ is a small positive constant. 
\subsubsection{Desired Signal Variance}
\par $\gamma_{nk}$ is related to the PSD of the desired signal $d_1[n,k]$. However, estimation of $\hat{{d}}_1[n,k]$ requires an estimate of $\bld{\gamma}_{nk}$. To avoid this dependence, the a-priori desired signal estimate
\begin{equation}\label{eqn:apriorisignal}
\tilde{d}_1[n,k] = \hat{\bld{w}}_{n-1}[k]^H \left( \bld{x}[n,k] - \hat{\bld{G}}_{n-1}[k]^H \bs{\phi}[n,k] \right),
\end{equation}
is used to compute the time domain AR model \eqref{eqn:armodelest} and the desired signal PSD estimate is obtained using \eqref{eqn:gammaest}.
\subsubsection{Relative transfer function $(\bld{a}_n)$} 
\par The time dependent RTF is computed using the estimated early reflection component $\hat{\bld{d}}[n]$. The spatial cross correlation matrix, similar to \eqref{eqn:predrescor}, is computed using recursive averaging with a smoothing parameter $\alpha_2$,
\begin{eqnarray}
\bld{R}_{\hat{\bld{d}}\hat{\bld{d}}}[n] = (1-\alpha_2) \bld{R}_{\hat{\bld{d}}\hat{\bld{d}}} [n-1] + \alpha_2 \hat{\bld{d}}[n] \hat{\bld{d}}[n]^H.
\end{eqnarray}
The estimate for RTF is computed similar to \eqref{eqn:hest}
\begin{equation}\label{eqn:rtfUpdate}
	\hat{\bld{a}}_n = \frac{ \bld{R}_{\hat{\bld{d}}\hat{\bld{d}}}[n] \bld{e}_1 }{ \bld{e}_1^H \bld{R}_{\hat{\bld{d}}\hat{\bld{d}}}[n] \bld{e}_1 }.
\end{equation}
Care must be taken in updating the RTF, since STFT bins with silence or dominant diffuse component in the estimated early reflection component can drift the RTF from the actual signal source. 
We determine the RTF update based on ratio of the average energies of $\hat{\bld{d}}[n]$ and $\hat{\bld{r}}[n]$, when it is greater than $0.1$. 
\subsection{Online RTF-MCLP Algorithm}
The algorithm for online RTF-MCLP is shown in Alg. \ref{alg:jointSFMCLP}. The spatial filter, and the MCLP prediction filter computed at the time index $n-1$ are used to obtain the a-priori desired signal estimate, which is then used to compute the desired signal variance. Given the desired signal variance, the posterior distribution parameters of the MCLP filters are computed and estimates of the multi channel early component and late reverberation component are computed. The estimated early reflection component is then used to update the RTF, and the late reverb component is used to update the noise covariance matrix. The spatial filter is then recomputed and the filter output is taken as the desired signal. 
\begin{algorithm}[h]
	\caption{Online joint spatial filtering and MCLP}\label{alg:jointSFMCLP}
	\begin{algorithmic}[1]
		\State Initialize $\hat{\bld{w}}_{-1}[k] = \bld{1}/\sqrt{M},~\forall~k$, $\{\bs{\mu}_{m,-1|-1}[k] = \bld{0},~\forall m,k\}$, $\bs{\Sigma}_{-1|-1}[k] = \eta \bld{I}$, $\eta=10^{-3}$. 
		\For {$n=0:N-1$}	

			\State Compute the a-priori signal estimate using \eqref{eqn:apriorisignal} $\forall k$.
			\State Compute the desired signal variance $\hat{\gamma}_{nk}$ using \eqref{eqn:gammaest}.
			
			\For {$k=0:K/2$}
			\State Compute $\Probability{\bld{g}_{m,n}[k] \given \bld{X}_{0:n-1}[k]},\forall~m$ using \eqref{eqn:kalmanPrior}.
			\State Compute $\Probability{ \bld{g}_{m,n}[k] \given \bld{X}_{0:n}[k] },\forall~m$ using \eqref{eqn:kalmanPost}.
			\State Update the prediction filter using \eqref{eqn:filtest}.
			\State Compute $\hat{\bld{d}}[n,k]$, and $\hat{\bld{r}}[n,k]$ using \eqref{eqn:sigest}.
			\State Update the RTF $\hat{\bld{a}}_n[k]$ using \eqref{eqn:rtfUpdate}.
			\State Update the noise covariance matrix using \eqref{eqn:noisecovarupdate}.
			\State Compute the spatial filter $\hat{\bld{w}}_n[k]$ using \eqref{eqn:mvdrSolution}.
			\State Compute the spatial filtered signal
				\begin{equation}
					\hat{{d}}_1[n,k] = \hat{\bld{w}}_n[k]^H \hat{\bld{d}}[n,k]
				\end{equation}			
			\EndFor							
		\EndFor		
	\end{algorithmic}
\end{algorithm}
\section{Experiments and Results}
\label{sec:experimentsAndResults}
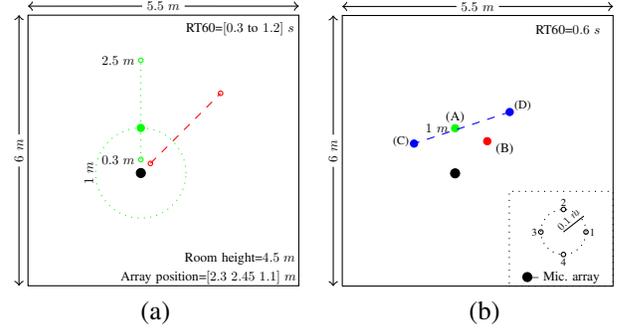
\begin{figure}[h]
	\centering
	\begin{minipage}{0.48\linewidth}
		\centering
		\begin{tikzpicture}[scale=0.6, every node/.style={scale=0.5}]
		\draw (0,0)--(6,0)--(6,6)--(0,6)--(0,0);	
		\node[rotate=90] at (-0.2,3) (A) {$6~m$};
		\draw[->] (A) -- (-0.2,6);
		\draw[->] (A) -- (-0.2,0);
		\node[] at (3,6.2) (B) {$5.5~m$};
		\draw[->] (B) -- (6,6.2);
		\draw[->] (B) -- (0,6.2);		
		\draw[fill=black] (2.5,2.5) circle (0.1);
		\draw[fill=green,green] (2.5,3.5) circle (0.08);
		\draw[green] (2.5,5) circle (0.05);
		\draw[green] (2.5,2.8) circle (0.05);		
		\draw[green,dotted] (2.5,2.8) -- (2.5,5.0);		
		\draw[red,thin] (2.7121,2.7121) circle (0.05); 
		\draw[red,thin] (4.2678,4.2678) circle (0.05); 
		\draw[red,dashed] (2.7121,2.7121) -- (4.2678,4.2678);		
		\node[] at (2.0,2.8) {$0.3~m$};
		\node[] at (2.0,5.0) {$2.5~m$};		
		\draw[green,dotted] (2.5,2.5) circle (1);
		\node[rotate=90] at (1.35,2.5){ $1~m$};	
		\node[] at (4.7,5.7) { RT60=$[0.3~\mbox{to}~1.2]~s$};
		\node[] at (4.7,.6) { Room height=$4.5~m$};
		\node[] at (4.0,.2) { Array position=$[2.3~2.45~1.1]~m$};		
		\end{tikzpicture}
		\centerline{(a)}
	\end{minipage}
	\begin{minipage}{0.48\linewidth}
		\begin{tikzpicture}[scale=0.6, every node/.style={scale=0.5}]
		\draw (0,0)--(6,0)--(6,6)--(0,6)--(0,0);	
		\node[rotate=90] at (-0.2,3) (A) {$6~m$};
		\draw[->] (A) -- (-0.2,6);
		\draw[->] (A) -- (-0.2,0);
		\node[] at (3,6.2) (B) {$5.5~m$};
		\draw[->] (B) -- (6,6.2);
		\draw[->] (B) -- (0,6.2);
		\draw[fill=black] (2.5,2.5) circle (0.1);
		\draw[fill=green,green] (2.5,3.5) circle (0.08);		
		\draw[red,fill=red] (3.2121,3.2121) circle (0.08); 
		\node[rotate=0] at (2.1,3.5){$1~m$};	
		\node[rotate=0] at (2.5,3.75){(A)};	
		\node[rotate=0] at (3.6,3.05){(B)};				
		
		\draw[fill=blue,blue] (3.71,3.86) circle (0.08);	
		\draw[fill=blue,blue] (1.59,3.16) circle (0.08);					
		\node[rotate=0] at (4,4){\small (D)};
		\node[rotate=0] at (1.3,3.2){\small (C)};
		
		\draw[-,dashed,blue] (3.71,3.86) -- (1.59,3.16);
		
		\node[] at (5.0,5.7) {RT60=$0.6~s$};
		
		\draw[dotted] (4.9,1.2) circle (0.5);
		\draw[] (4.9,1.2) --node[left,above,rotate=45,scale=0.8]{ $0.1~m$} (5+0.707/2,1.2+0.707/2);
		\draw[] (5.4,1.2) circle (0.1/2) node[right,scale=0.8] { 1};
		\draw[] (4.9,1.7) circle (0.1/2) node[above,scale=0.8] { 2};
		\draw[] (4.4,1.2) circle (0.1/2) node[left,scale=0.8] { 3};
		\draw[] (4.9,0.7) circle (0.1/2) node[below,scale=0.8] { 4};
		\draw[fill=black] (4.1,0.2) circle (0.1) node[right] { -- Mic. array};
		\draw[dotted] (3.7,0) rectangle (6,2.1);
		\end{tikzpicture}
		\centerline{(b)}		
	\end{minipage}
	\caption{Room setup for simulated impulse responses. (a) Stationary source case, (b) Dynamic source case. Black: UCA mics located at $[2.3,2.45,1.1]~m$ for both cases.}\label{fig:simulationRoomSetup}
\end{figure}

First we study the performance of the joint RTF-MCLP model for signal estimation in a batch mode (static sources but non-stationary speech), and later present the results for a dynamic source. We consider the recording setup shown in Fig. \ref{fig:simulationRoomSetup}; we generate simulated room impulse responses, using the image method \cite{allen1979image, habets2010room}, for each of the mics. A uniform circular array (UCA) with $4$ microphones and of radius $0.1~m$ is considered for the recording. Center of the array is placed at coordinates $[2.3,2.45,1.1]~m$ with respect to the lower left corner of the enclosure. Desired source is assumed to be at the position $[2.3,3.45,1.1]~m$, and an interferer at $[3.007,3.157,1.1]~m$ (both at a distance of $1~m$ from the center of array), and the reverberation time of the enclosure is $RT60=0.6~s$, unless otherwise stated.
\par Clean signals from the TIMIT database \cite{timit} are convolved with the simulated RIRs to generate the reverberant microphone signals. Ten signals ($5$ each from male and female speakers) are used for the evaluation. Interference signal is also speech signal from different from the target signal. STFT analysis of mic signal is carried out using a Hann window of size $32~ms$, and a successive frame overlap of $75\%$. We consider the size of discrete Fourier transform in STFT to be the same as window size (equal to $512$ for the signal sampling rate of $16~KHz$).
\par We study the performance of the proposed dereverberation approach using the objective measure of frequency weighted signal-to-noise ratio (FwSNR) \cite{Hu2008Evaluation}, and subjective performance measures of ``perceptual evaluation of speech quality'' (PESQ) \cite{Rix2001Perceptual} and the ``short-time objective intelligibility'' (STOI) \cite{taal2010short}. PESQ is bounded between $[1-5]$ with $5$ being closest to clean speech. STOI lies between $[0-1]$ with $1$ indicating best intelligibility. FwSNR is computed using the evaluation toolbox provided in \cite{Kinoshita2016Summary}, and STOI is computed using the auditory modeling toolbox \cite{SoendergaardMajdak2013}. Processed speech examples are available online at {\url{www.ece.iisc.ernet.in/~sraj/dMCLP.html}}.
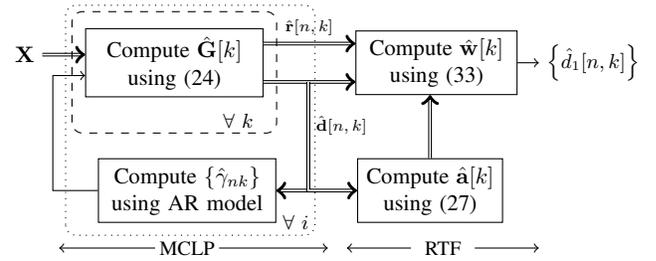
\begin{figure}[h]
	\centering
	\begin{tikzpicture}[scale=0.85, every node/.style={scale=0.85}]
	\node[] (inNode) at (-1.0in,0.05in) {${\bf X}$}; 
	\node[draw,text width=1in,align=center] (filter) at (-0.2,0) {Compute $\hat{{\bf G}}[k]$ using \eqref{eqn:gest}};		
	\node[draw,text width=0.8in,align=center] (filtest) at (3.8,-2.0) {Compute $\hat{\bld{a}}[k]$ using \eqref{eqn:hest} };		
	\node[draw,text width=0.9in,align=center] (residual) at (3.9,0) {Compute $\hat{\bld{w}}[k]$ using \eqref{eqn:west2}};		
	
	\draw[double,->] ([yshift=3.0mm] filter.east) --node[above,pos=0.5]{\scriptsize $\hat{\bld{r}}[n,k]$} ([yshift=3.0mm] residual.west);
	
	\draw[double,->] (filtest) -- (filtest |- residual.south);
	\node[draw,text width=1.0in,align=center] (psd) at (0,-2.0) {Compute $\{\hat{\gamma}_{nk}\}$ using AR model};
	\draw[double,->] ([yshift=-3mm]filter.east) -- ([yshift=-3mm]residual.west);
	
	\draw[->,double] ([xshift=7mm,yshift=-3mm]filter.east) |-node[above,right,pos=0.2]{\scriptsize $\hat{\bld{d}}[n,k]$} (psd.east);
	\draw[->,double] ([xshift=7mm,yshift=-3mm]filter.east) |- (filtest.west);
	
	\draw[->] (psd.west) -- (-2.1,-2.0) |- ([yshift=-2mm]filter.west);
	\draw[double,->] (inNode) -- (inNode -| filter.west);
	\node[] (outNode) at (6.35,0) {\small $\inCurly{\hat{d}_1[n,k]}$};
	
	\draw[dashed, rounded corners] (-1.8,-1.1) rectangle (1.4,0.8);
	\draw[dotted, rounded corners] (-1.9,-2.7) rectangle (2,0.9);
	\node[] at (1.7,-2.5) {\small $\forall~i$};
	\node[] at (0.8,-0.9) {\small $\forall~k$};
	\draw[->] (residual.east) -- (outNode);
	
	\draw[<-] (-2,-2.9) -- (-0.5,-2.9);\node[ ]  at (0,-2.9) {\small MCLP};\draw[->] (0.5,-2.9) -- (2.2,-2.9);
	\draw[<-] (2.5,-2.9) -- (3.5,-2.9);\node[ ]  at (4,-2.9) {\small RTF};\draw[->] (4.5,-2.9) -- (5.5,-2.9);
	\end{tikzpicture}	
	\vspace{-7pt}
	\caption{Cascaded MCLP and RTF based spatial filtering scheme \cite{cohen2017combined, delcroix2014linear} for performance comparison with RTF-MCLP.}\label{fig:srtfmclp}
\end{figure}
\par We compare the performance of the new RTF-MCLP with (i) MCLP method with AR source prior \cite{integrated2009yoshioka} (first block in Fig.\ref{fig:srtfmclp}), (ii) a separate cascade of the MVDR beamformer for the MCLP output \cite{delcroix2014linear, cohen2017combined} (C-RTF-MCLP) as shown in Fig.\ref{fig:srtfmclp}, and (iii) the super directive beamformer (SDB) \cite{capon1969high, brandstein2013microphone}. For the MCLP method, the autocorrelation function required for source AR estimation is computed by averaging the autocorrelation function of the multi-channel prediction residual signals. For the MVDR beamformer post filter, the noise covariance matrix is computed using the reverberation component estimated using the MCLP, and the RTF is estimated using \eqref{eqn:hest}. SDB is implemented assuming free field RTF $\bld{a}$ (vector $\bld{a}$ corresponds to the direct path only) and the microphone placement (diffuse noise spatial covariance matrix) is assumed known a-priori.
\subsection{Stationary source, batch processing}
	\vspace{-5pt}
\begin{figure}[h]
	\centering
	\begin{minipage}[b]{0.48\linewidth}
		\centerline{\includegraphics[width=1.6in,height=1.3in]{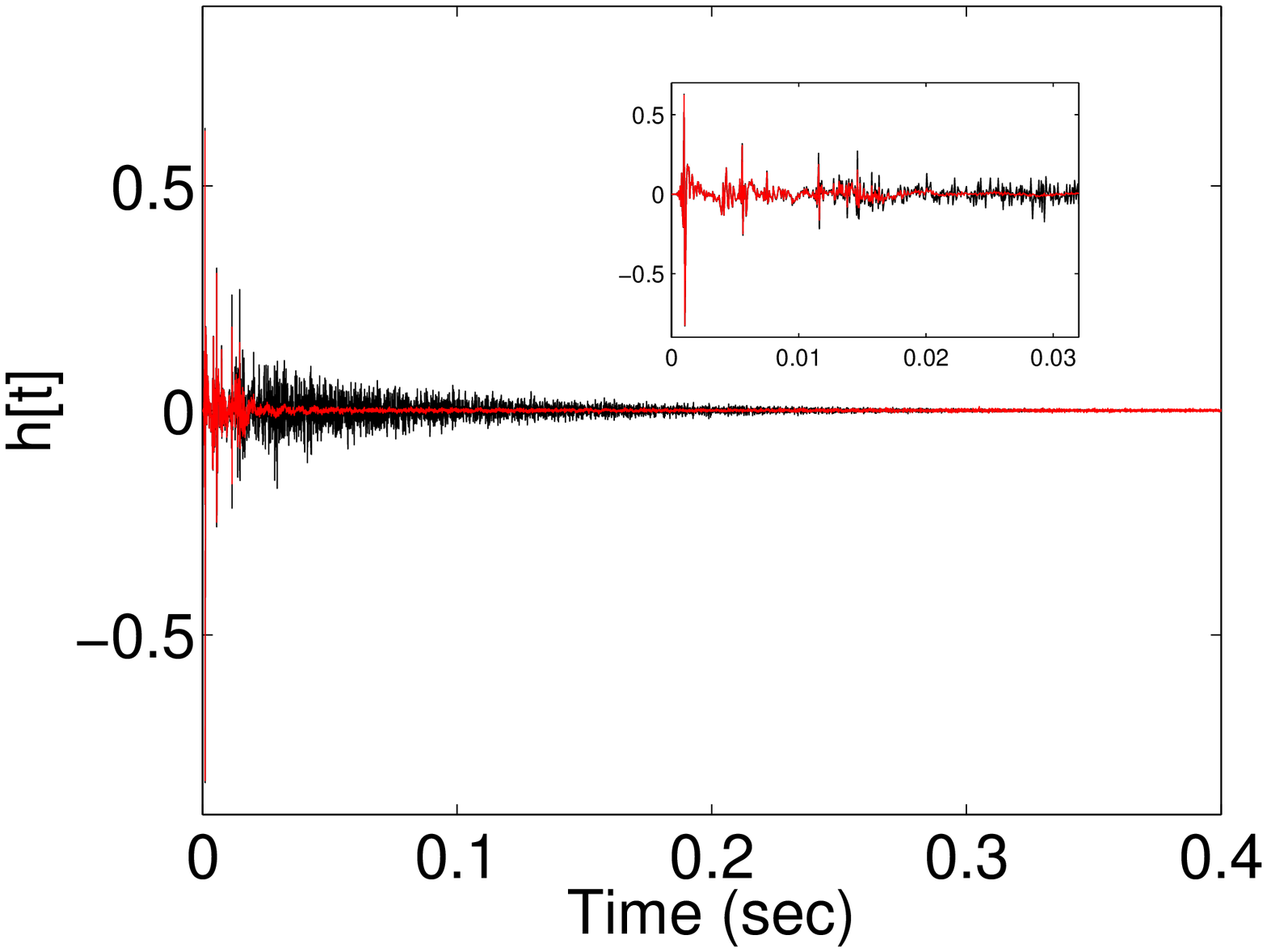}}
		\centerline{(a)}
	\end{minipage}	
	\begin{minipage}[b]{0.48\linewidth}
		\centerline{\includegraphics[width=1.6in,height=1.3in,trim={0pt 0pt 0pt 15pt}]{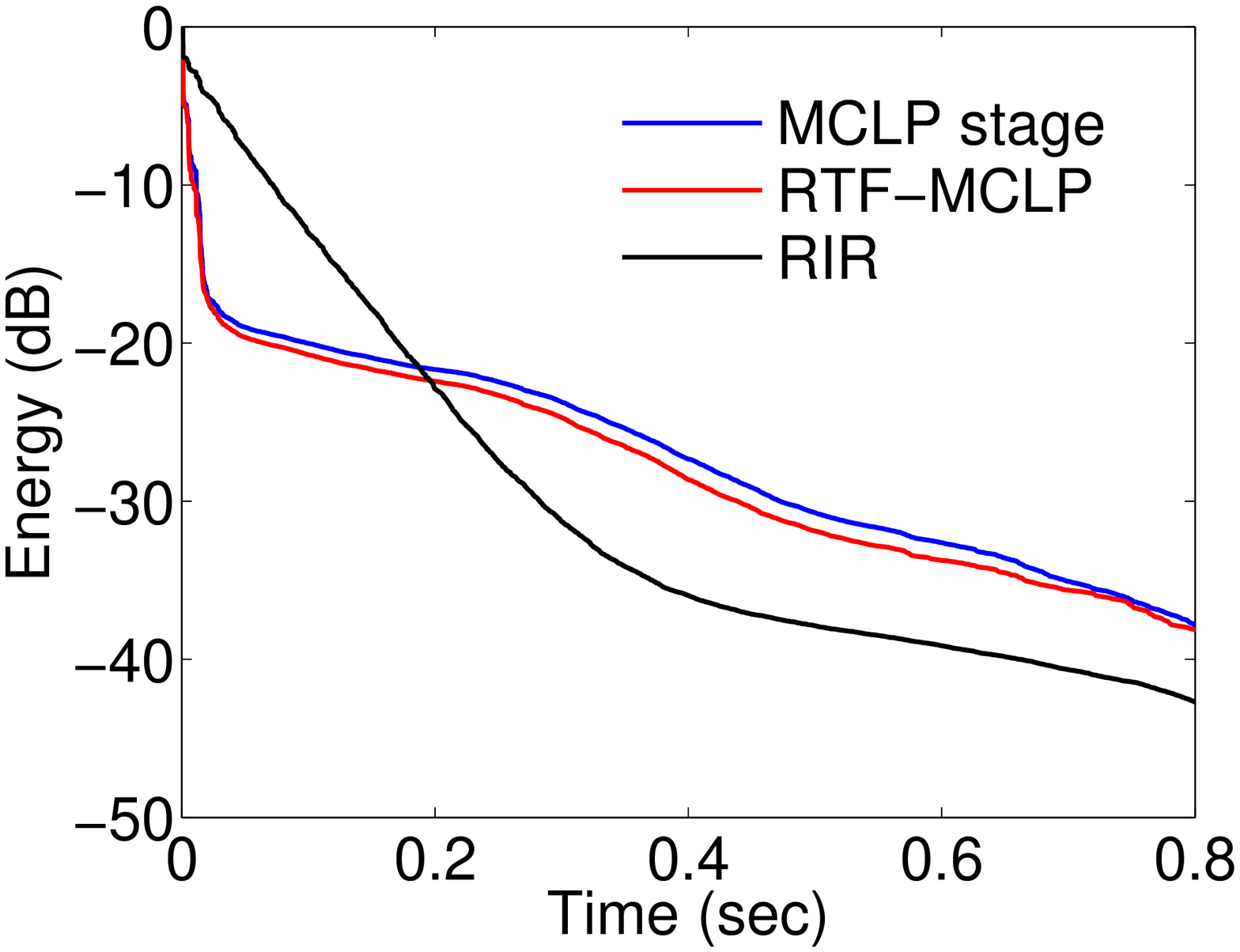}}
		\centerline{(b)}
	\end{minipage}
	\vspace{-5pt}
	\caption{Room impulse response illustration.}\label{fig:rirEdc}
\end{figure}
\par Fig. \ref{fig:rirEdc}(a) shows the output of the MCLP filters in RTF-MCLP scheme with RIR as the input, in time domain. The filters are estimated first using a speech signal, and then used to obtain the output RIR considering it as the microphone signal. We see from estimated RIR that the direct component and the first few early reflection peaks are retained; they also match the input RIR in amplitude and time-location. The reflection components beyond $\approx 16~ms$ are suppressed significantly after the MCLP filtering, i.e., the effective duration of RIR after the MCLP filter stage is $\approx 16~ms$. Since the window size chosen for the STFT analysis is $32~ms$, we can see that the RTF approximation in the STFT domain is valid, only with some residual reverberation noise. Fig. \ref{fig:rirEdc}(b) shows the corresponding energy decay curve (EDC), we can see that the room reflection components over long-term of $16-220~ms$ are suppressed. However, there is an increase in the reflection component energy beyond $\approx 220~ms$, but it is below $-20~dB$ level. Fig. \ref{fig:rirEdc}(b) also shows EDC for the effective RIR after the spatial filter. We see that the level of residual reverberation decreases after the spatial filter, and we found that there is a small improvement near the early reflection components also.
\subsubsection{RTF-MCLP Algorithm Parameters}
\par The order of the MCLP filter $L$ and the delay parameter $D$ are the parameters of the algorithm, and we explore each parameter individually by fixing the other. We fix the baseline experimental parameters as \{max iterations$=5$, $L=12$, $M=4$, $D=2$\}. The acoustic room setup is fixed as \{source distance$=1.0~m$, source angle$=90^o$, $RT60=0.6~s$\}, shown in Fig. \ref{fig:simulationRoomSetup}.
\begin{figure}[h]
	\centering
	\centerline{\includegraphics[width=3in,height=0.2in,trim={20pt 25pt 40pt 30pt},clip]{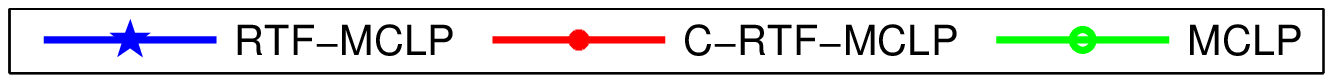}}	
	\begin{minipage}[b]{0.3\linewidth}
		\centerline{\includegraphics[width=1.in,height=1.in]{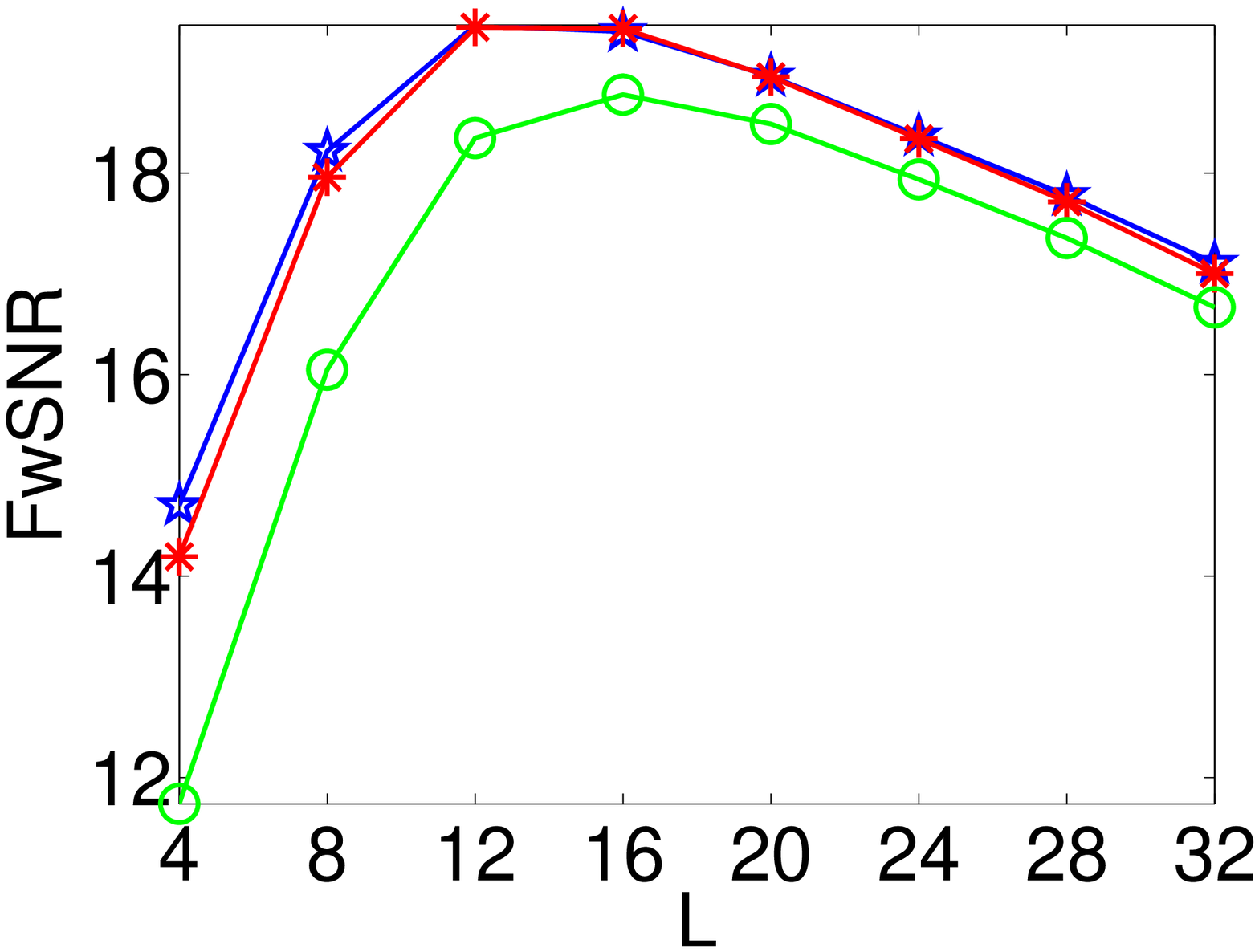}}
		\vspace{-2pt}
		\centerline{(a)}
	\end{minipage}
	\begin{minipage}[b]{0.3\linewidth}
		\centerline{\includegraphics[width=1in,height=1.in]{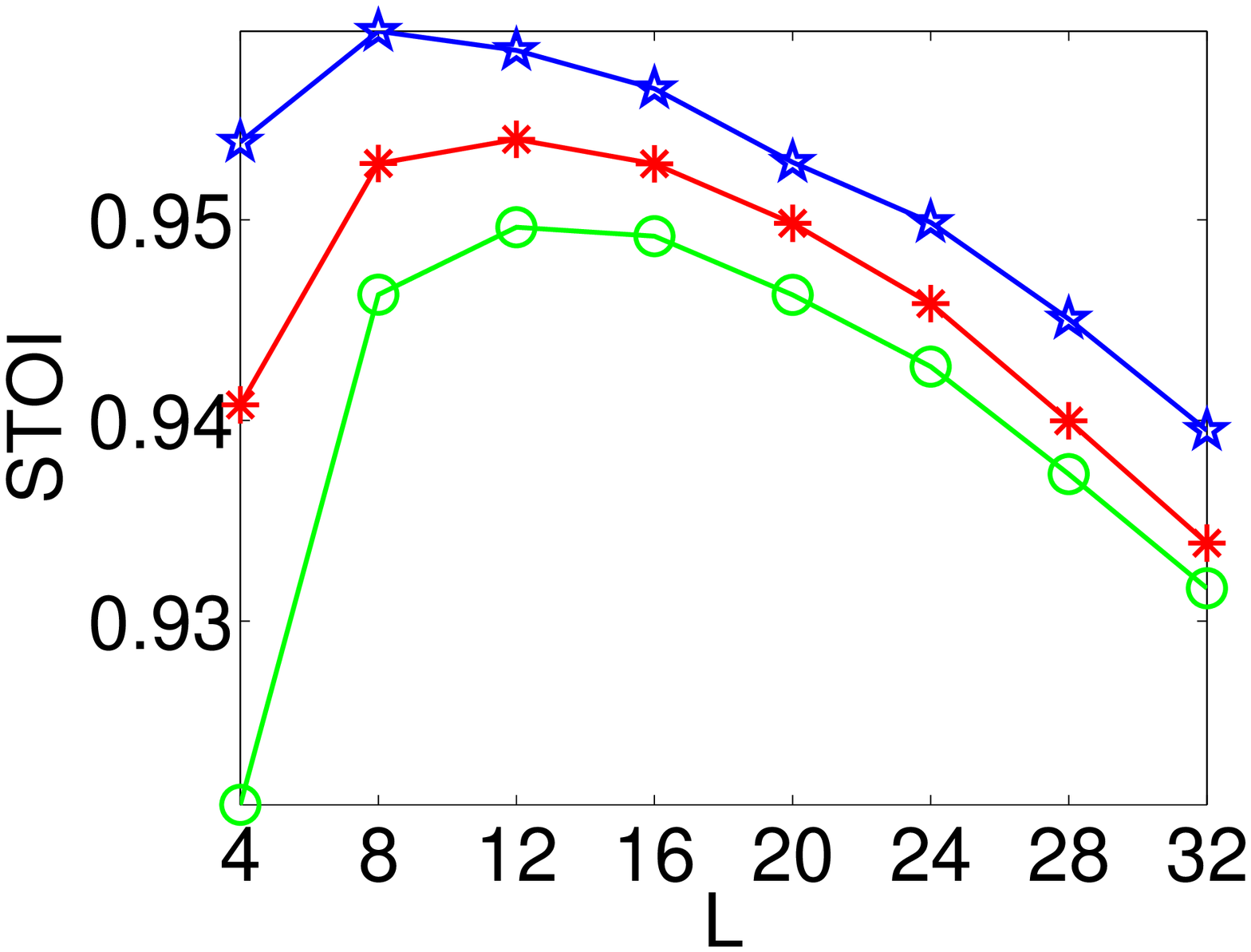}}
		\vspace{-2pt}
		\centerline{(b)}
	\end{minipage}
	\begin{minipage}[b]{0.3\linewidth}
		\centerline{\includegraphics[width=1.in,height=1.in]{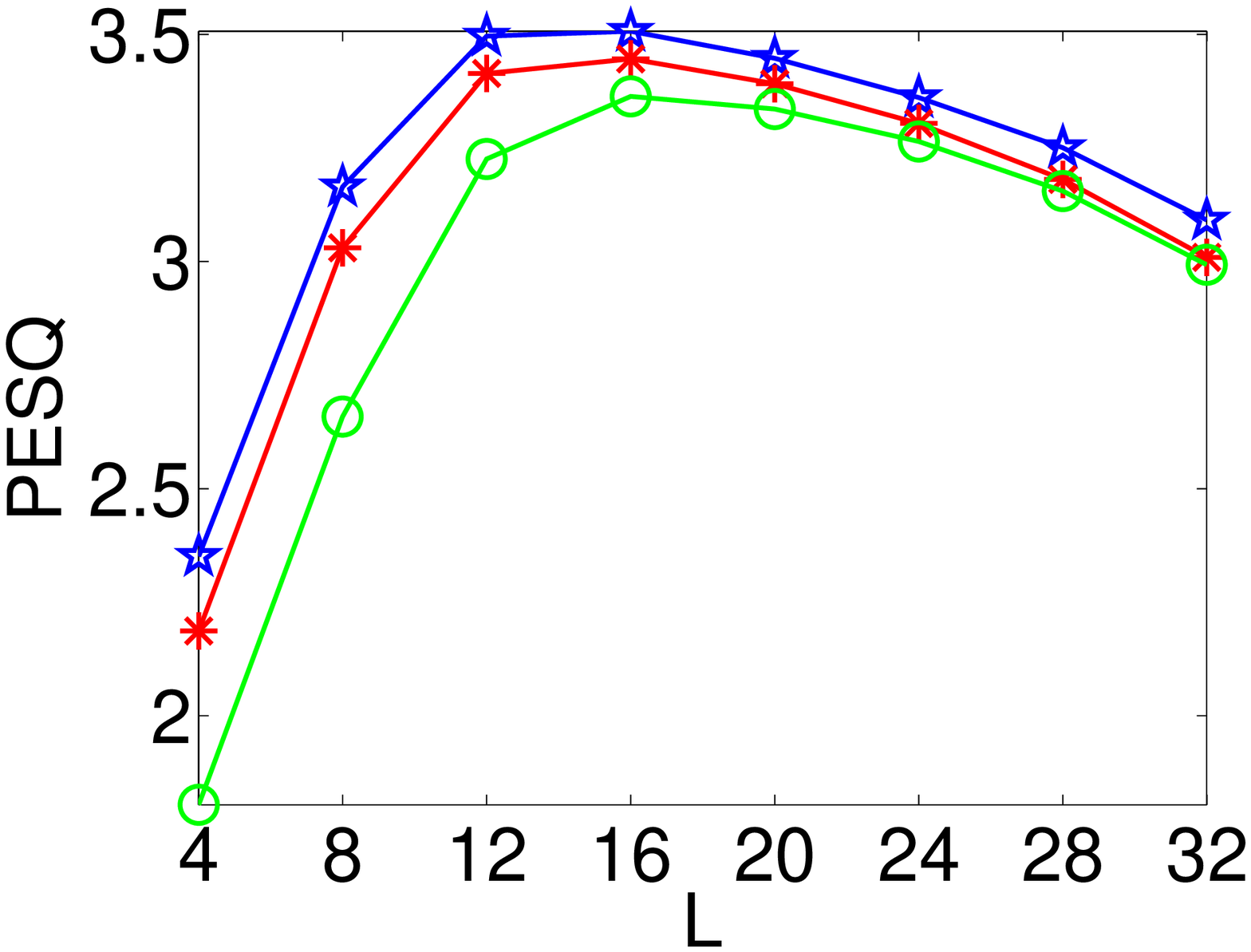}}
		\vspace{-2pt}
		\centerline{(c)}
	\end{minipage}
	
	\begin{minipage}[b]{0.3\linewidth}
		\centerline{\includegraphics[width=1.in,height=1.in]{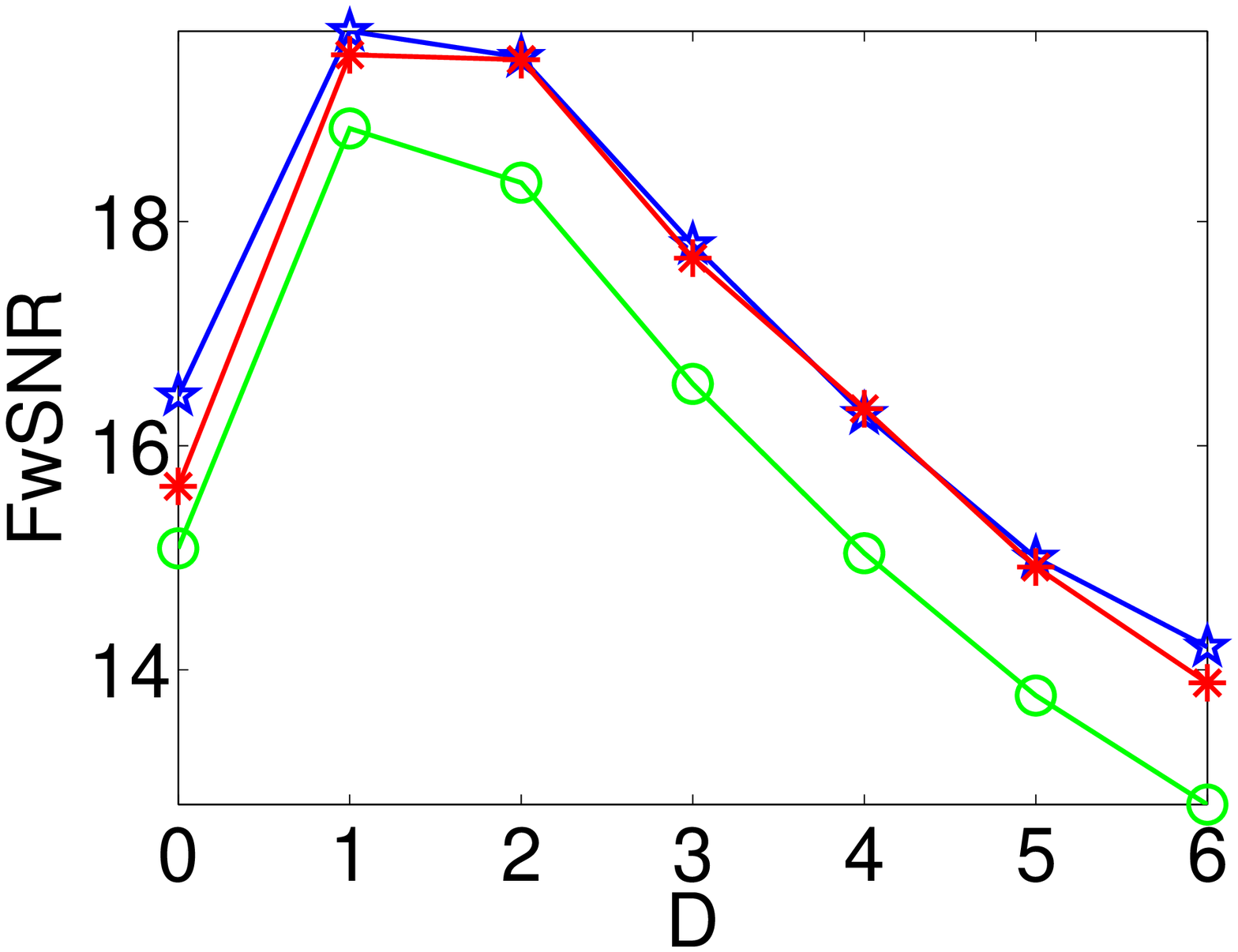}}
		\vspace{-2pt}
		\centerline{(d)}
	\end{minipage}
	\begin{minipage}[b]{0.3\linewidth}
		\centerline{\includegraphics[width=1.in,height=1.in]{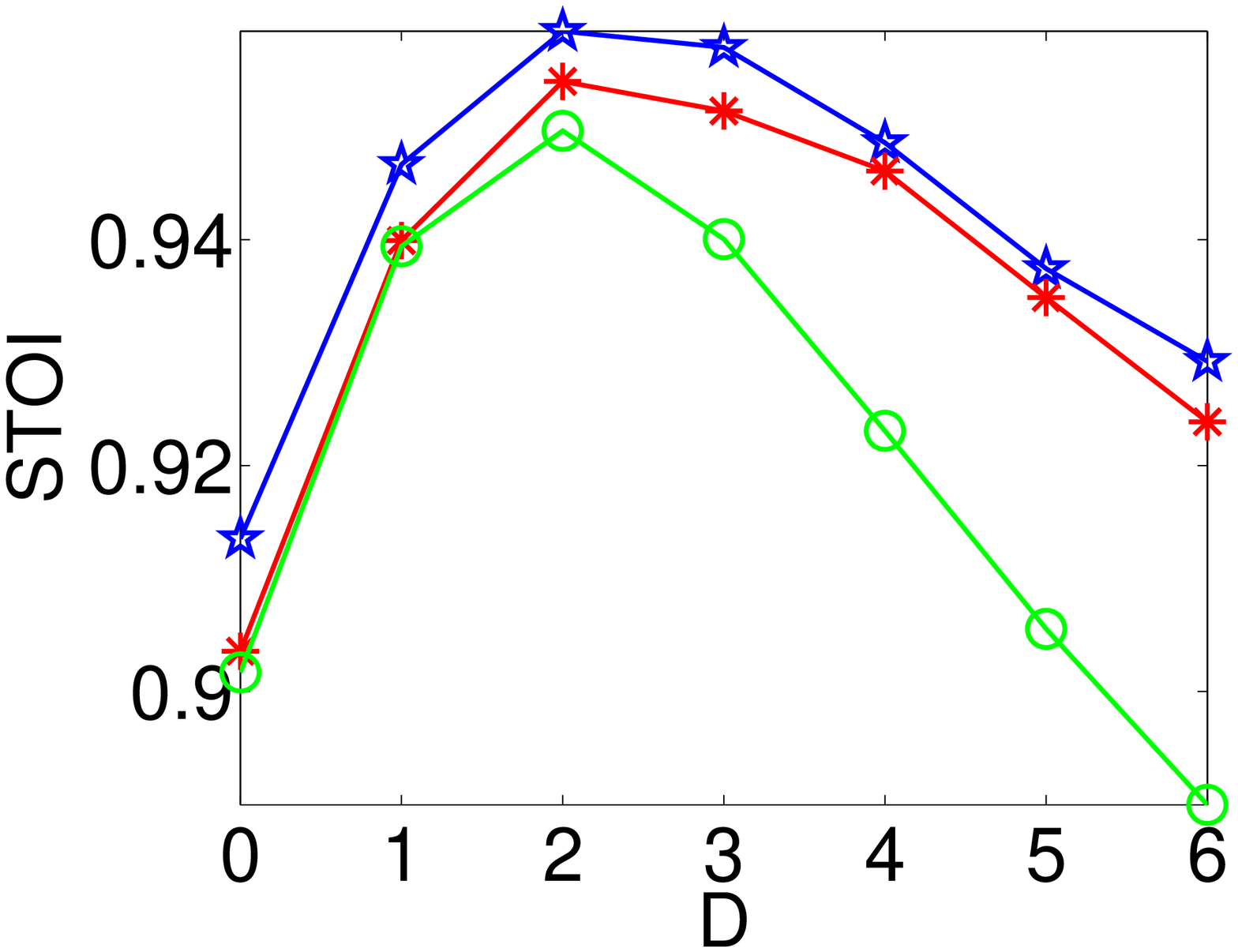}}
		\vspace{-2pt}
		\centerline{(e)}
	\end{minipage}
	\begin{minipage}[b]{0.3\linewidth}
		\centerline{\includegraphics[width=1.in,height=1.in]{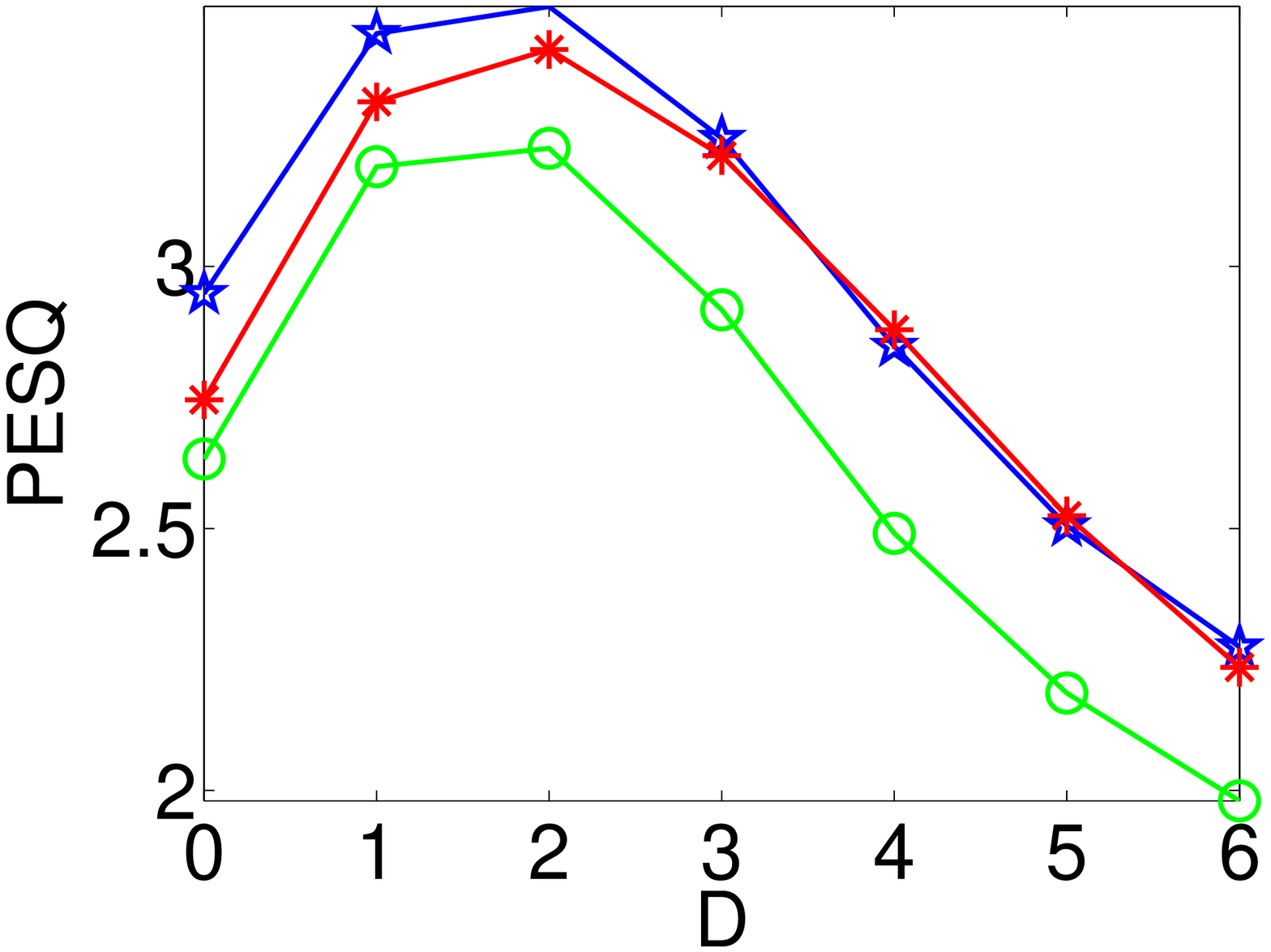}}
		\vspace{-2pt}
		\centerline{(f)}
	\end{minipage}				
	\caption{Dereverberation performance for different algorithm conditions. The rows show the results for MCLP Order ($L$), and delay parameter ($D$) variation respectively. The three columns show the performance measures average FwSNR, STOI and PESQ. Average FwSNR, STOI and PESQ for reverberant input signal are \{$7.37~dB$, $0.81$ and $1.38$\} respectively. For the super directive beamformer output $\{ 10.76~dB,~0.91,~\mbox{and}~1.63\}$ respectively. }\vspace{-5pt}
	\label{fig:performanceAlgorithmParameters}
\end{figure}
\par Fig. \ref{fig:performanceAlgorithmParameters} shows the performance measures for different algorithm parameters. We see that (i) spatial filter in combination with the MCLP filter improves the MCLP performance, and (ii) the joint estimation in RTF-MCLP performs better than the cascade approach of C-RTF-MCLP, particularly for the perceptual measures of PESQ and STOI, though minimal for FwSNR. 
Fig. \ref{fig:performanceAlgorithmParameters}(a-c) show the performance measures for increasing value of prediction order $L$. We see that peak performance is obtained for $L=12\mbox{ to }16$, indicating the structure of diffuse component in the STFT domain. Performance for different values of $D$ is shown in Fig. \ref{fig:performanceAlgorithmParameters}(d-f). Again there is a peak performance for $D=1$ and $D=2$. $D=0$ case uses immediate past value for prediction, which results in degradation of the estimated signal due to the high correlation between successive STFT frames. Higher values of $D$ retains more of the diffuse component in the desired signal, hence the degradation in performance compared to the clean signal. Also, for higher $D$, the RTF approximation does not hold good resulting in performance degradation. Yet the performance is found to be better than using only MCLP in all the cases. 
\subsubsection{RIR conditions}
\par RIR is a function of the distance and angular position of the source with respect to the microphones, and $RT60$ value of the enclosure. These parameters are studied individually by fixing the other two parameters. The source distance and the angular position are varied as shown in Fig. \ref{fig:simulationRoomSetup} (dotted line and circle in green), and the RT60 is varied from $0.3~s$ to $1.2~s$.
The baseline values for the parameters are \{source distance$=1.0~m$, source angle$=90^o$, $RT60=0.6~s$\}. The algorithm parameters are fixed as \{max iterations$=5$, $L=12$, $M=4$, $D=2$\}.
\par
\begin{figure}[h]
	\centering
	\centerline{\includegraphics[width=3in,height=0.2in,trim={40pt 60pt 90pt 22pt},clip]{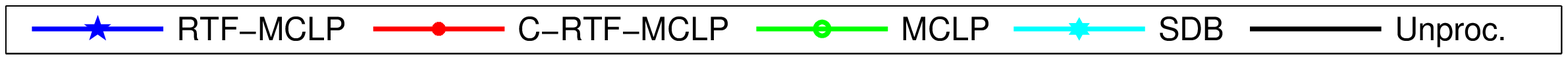}}	
	\begin{minipage}[b]{0.3\linewidth}
		\centerline{\includegraphics[width=1in,height=1.05in]{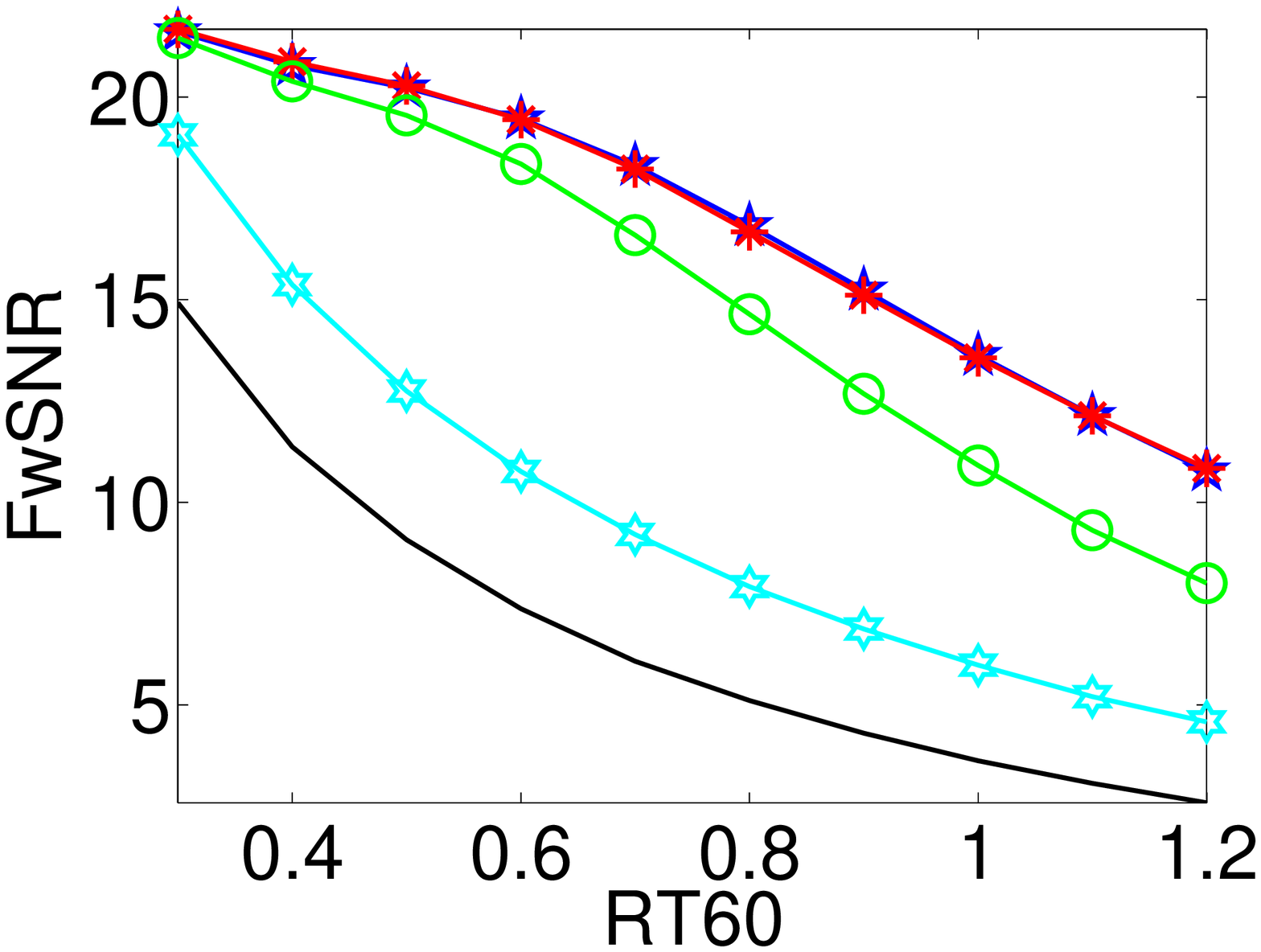}}
		\centerline{(a)}
	\end{minipage}
	\begin{minipage}[b]{0.3\linewidth}
		\centerline{\includegraphics[width=1in,height=1.05in]{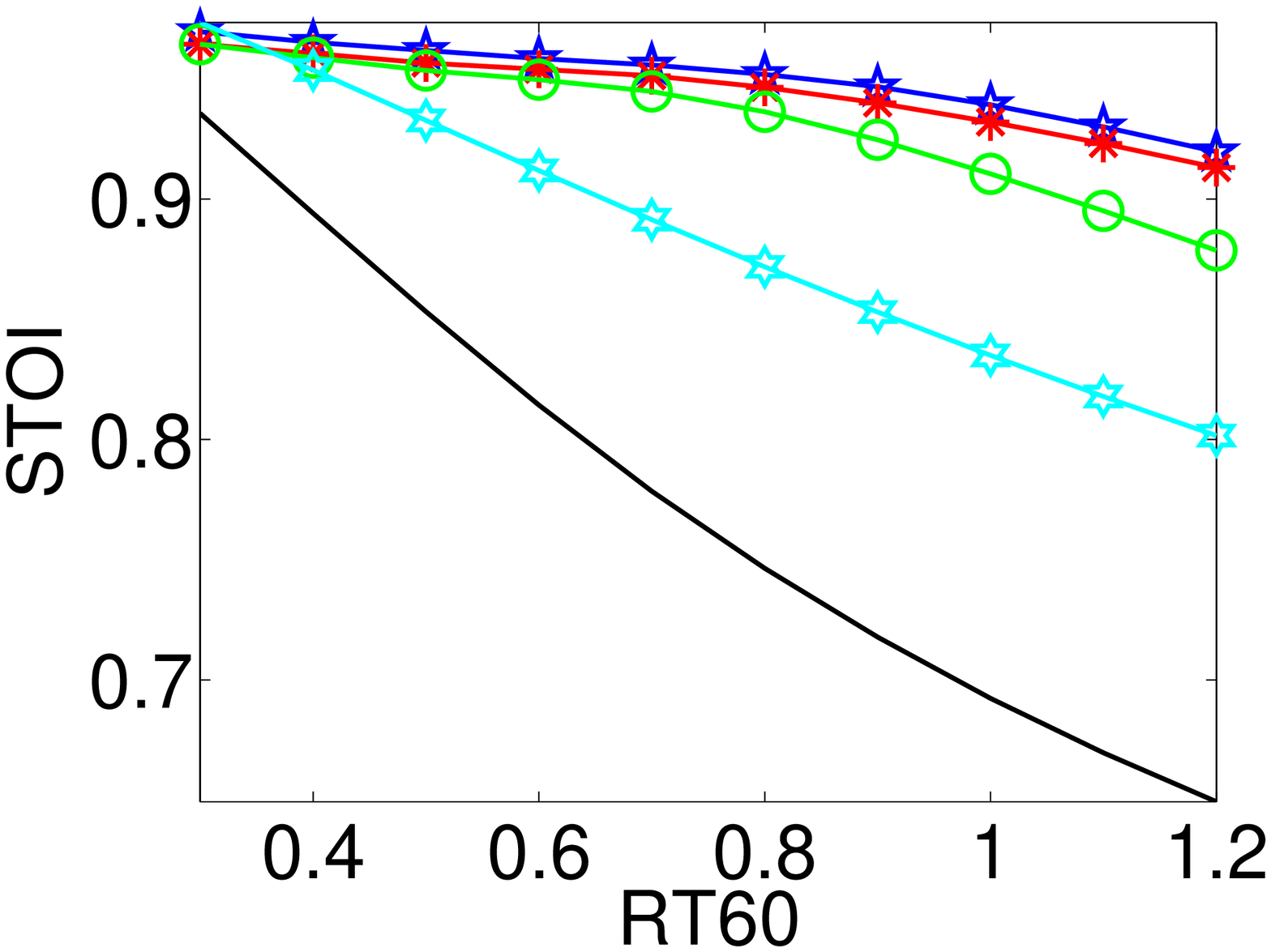}}
		\centerline{(b)}
	\end{minipage}
	\begin{minipage}[b]{0.3\linewidth}
		\centerline{\includegraphics[width=1in,height=1.05in]{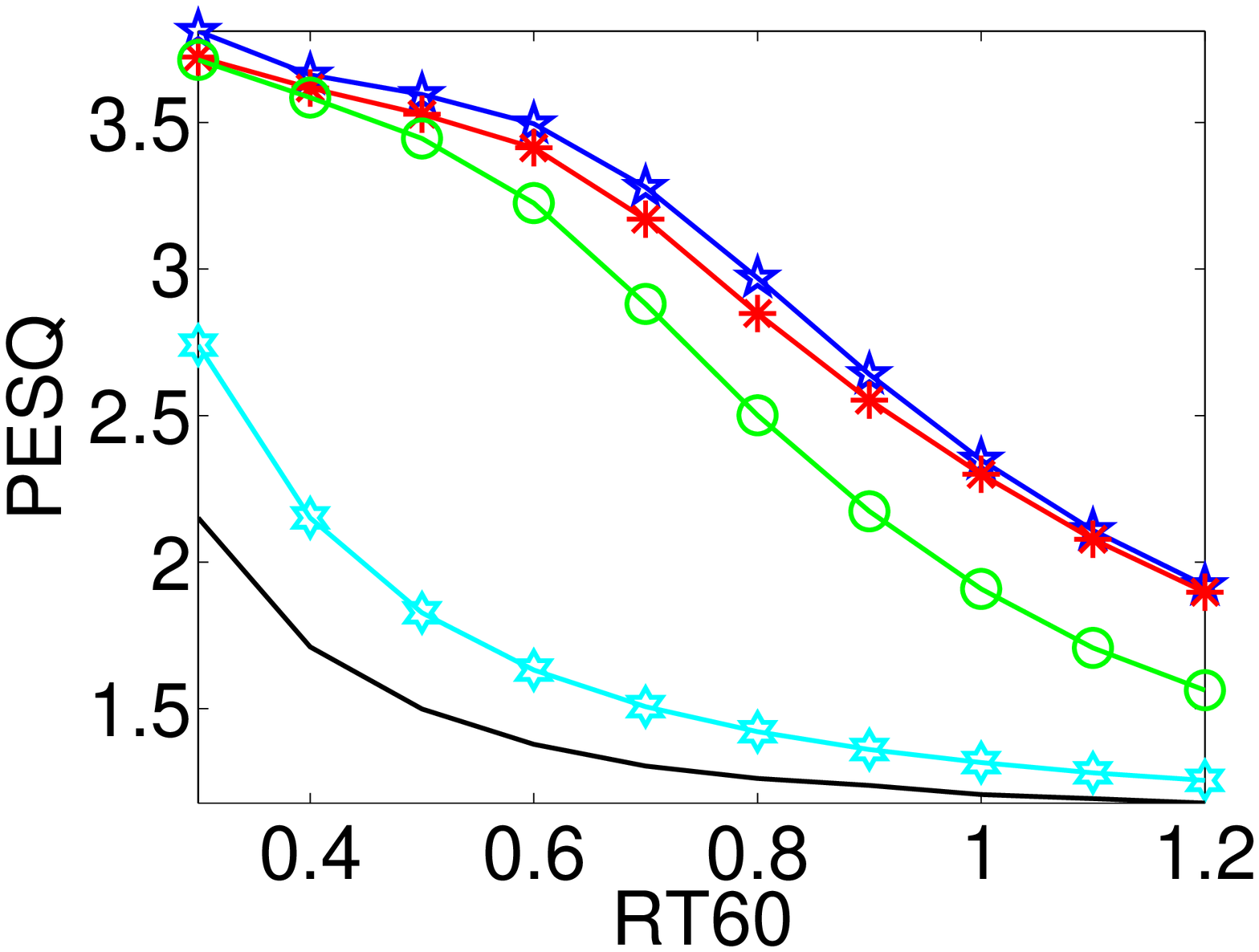}}
		\centerline{(c)}
	\end{minipage}

	\begin{minipage}[b]{0.3\linewidth}
		\centerline{\includegraphics[width=1in,height=1.05in]{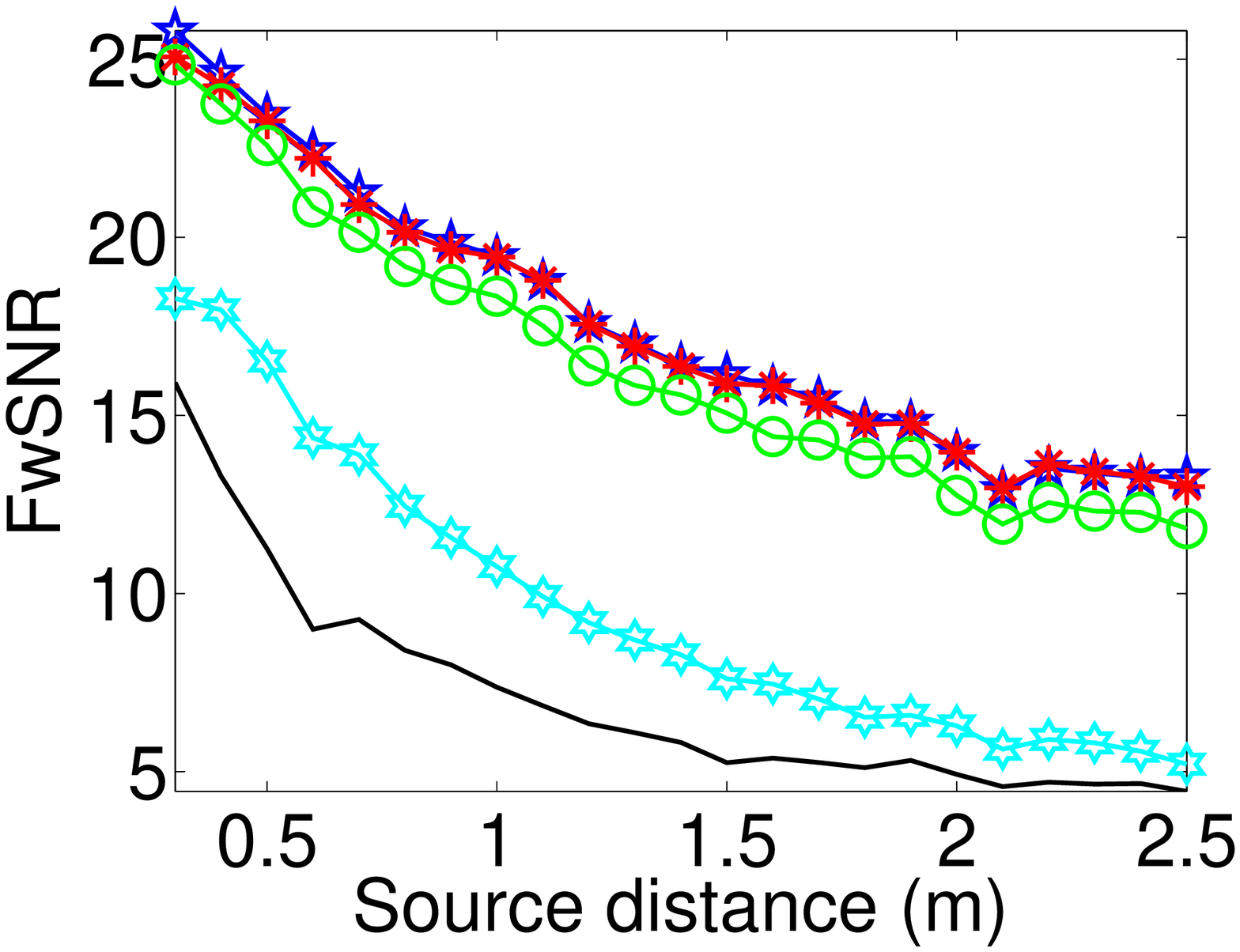}}
		\centerline{(d)}
	\end{minipage}
	\begin{minipage}[b]{0.3\linewidth}
		\centerline{\includegraphics[width=1in,height=1.05in]{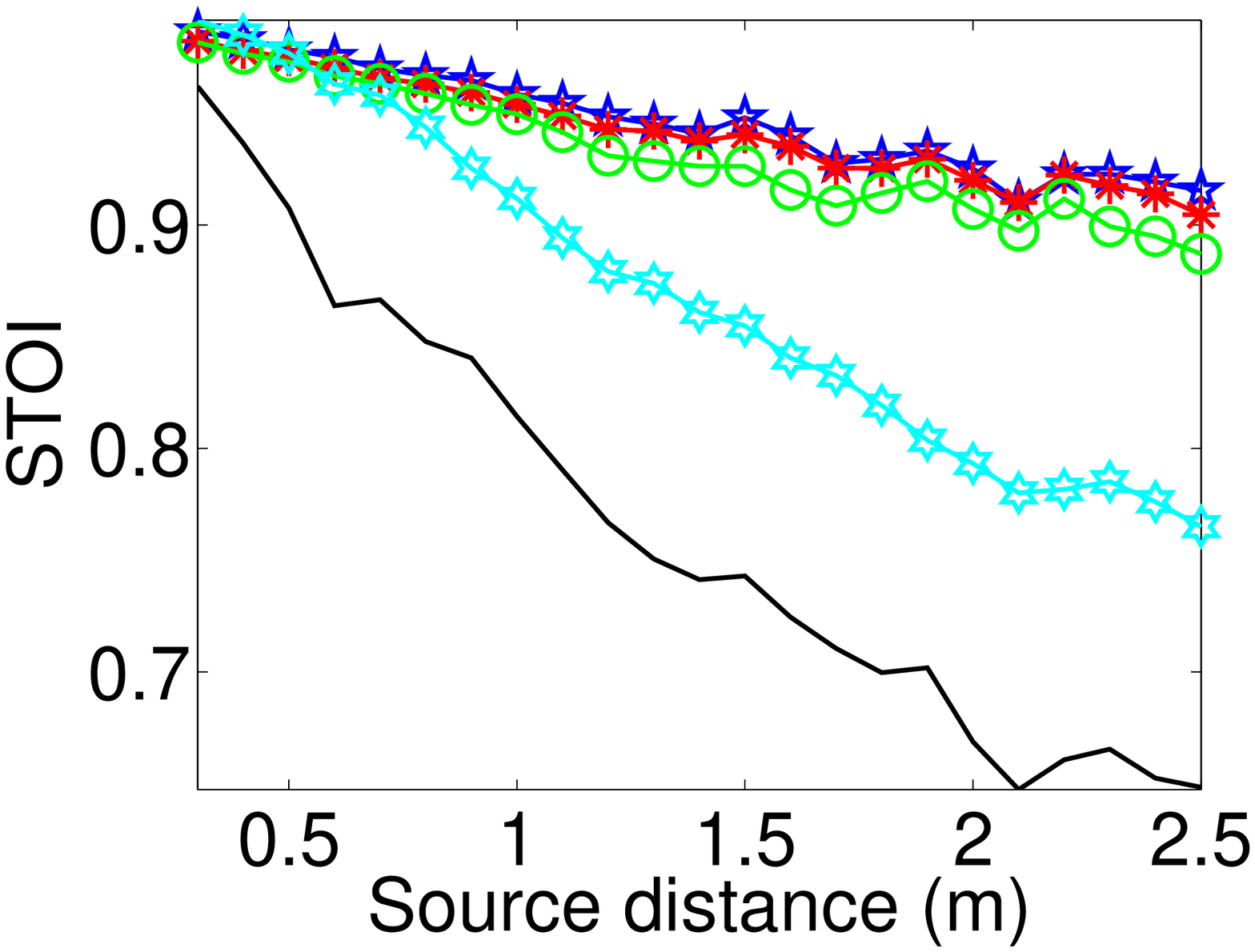}}
		\centerline{(e)}
	\end{minipage}
	\begin{minipage}[b]{0.3\linewidth}
		\centerline{\includegraphics[width=1in,height=1.05in]{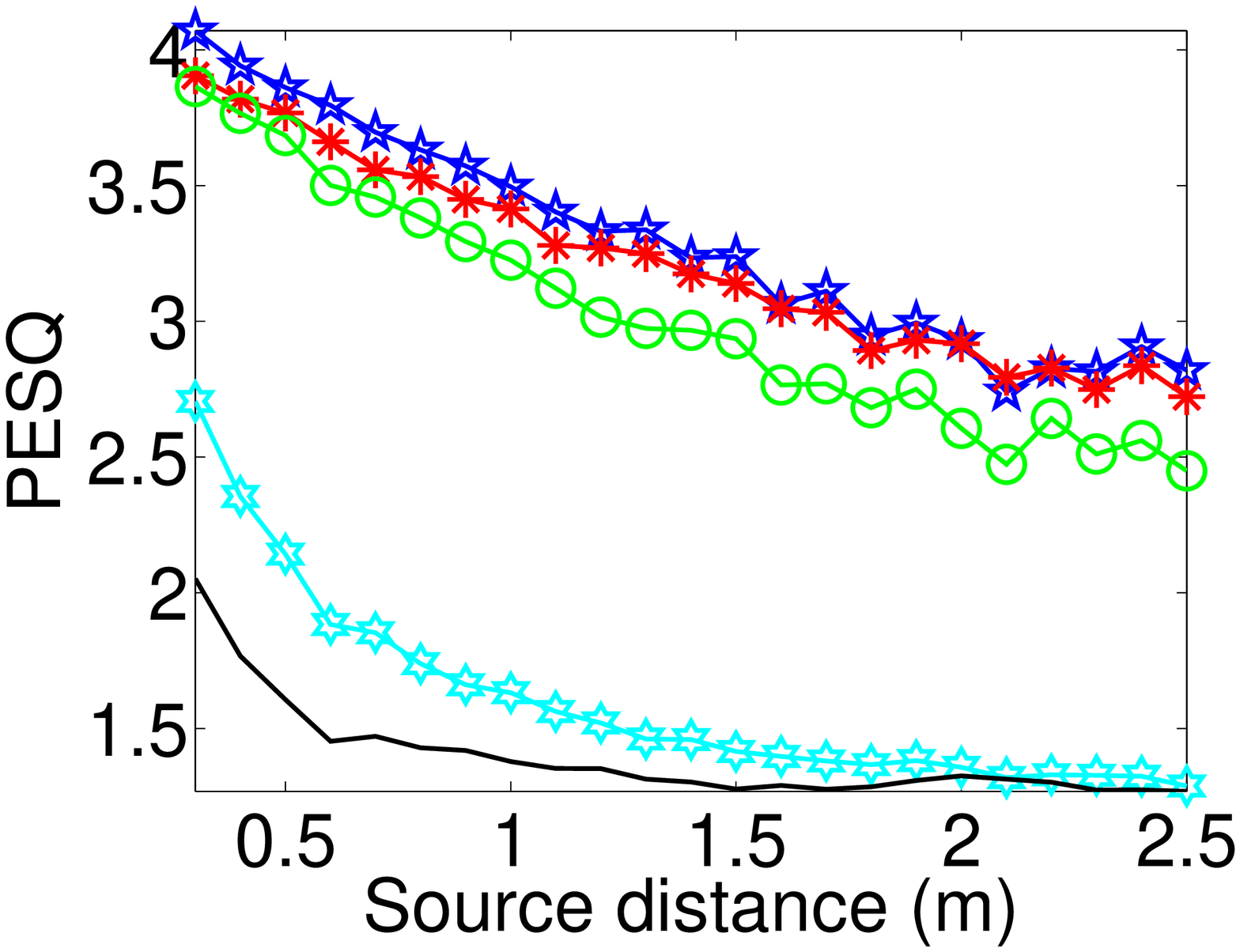}}
		\centerline{(f)}
	\end{minipage}
	
	\begin{minipage}[b]{0.3\linewidth}
		\centerline{\includegraphics[width=1in,height=1.05in]{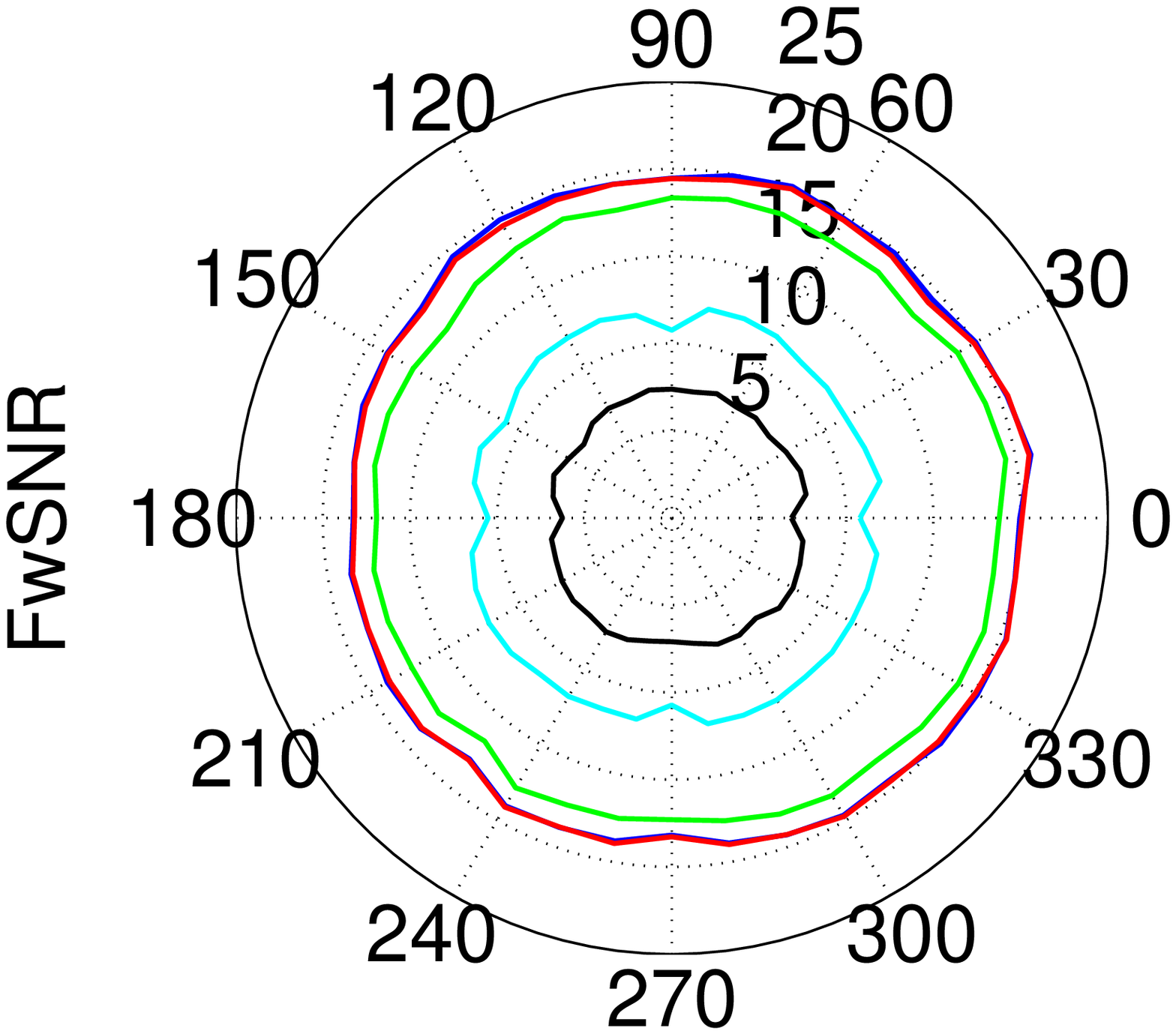}}
		\centerline{(g)}
	\end{minipage}
	\begin{minipage}[b]{0.3\linewidth}
		\centerline{\includegraphics[width=1in,height=1.05in]{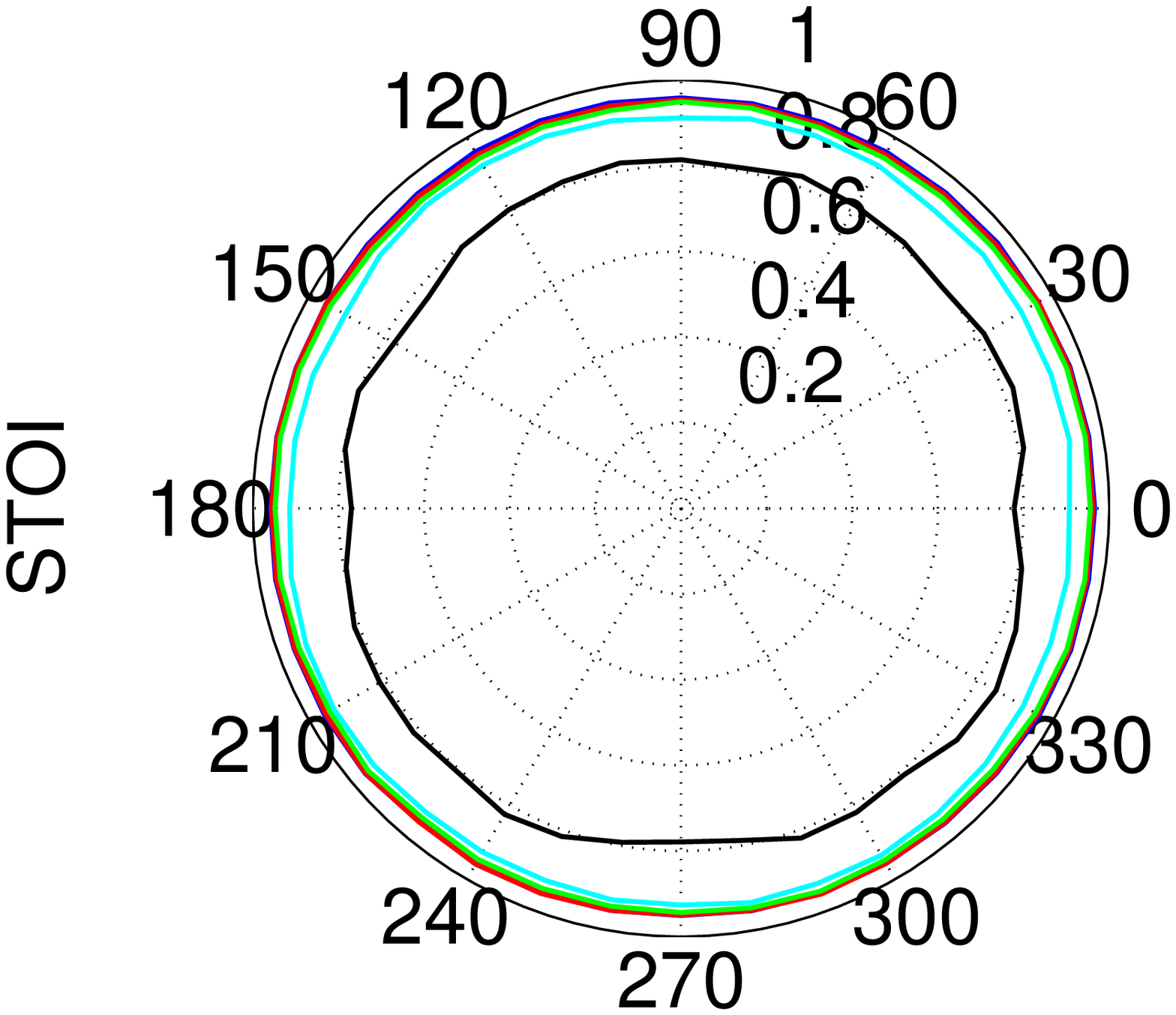}}
		\centerline{(h)}
	\end{minipage}
	\begin{minipage}[b]{0.3\linewidth}
		\centerline{\includegraphics[width=1in,height=1.05in]{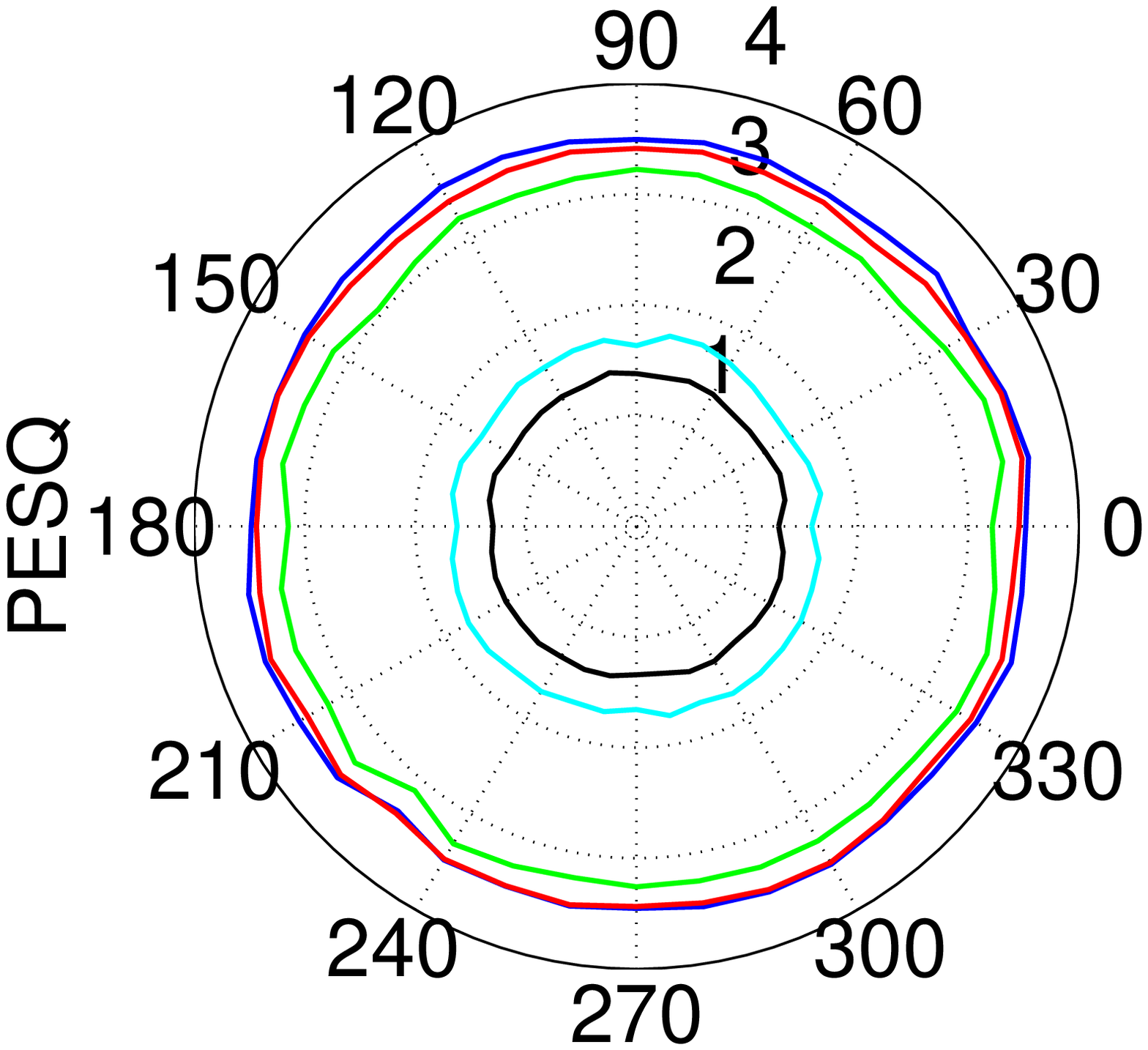}}
		\centerline{(i)}
	\end{minipage}
	\caption{Dereverberation performance for (a-c) changing $RT60$, (d-f) changing source distance, and (g-i) changing source direction. In (g-i) performance measures are mapped to radial distance. The three columns show the performance measures average FwSNR, STOI and PESQ.}\vspace{-5pt}
	\label{fig:performanceAcousticParameters}
\end{figure}

 Fig. \ref{fig:performanceAcousticParameters} shows the performance comparison for different RIR conditions. We can see that, (i) spatial filtering aids MCLP and hence improves the dereverberation performance of MCLP method, (ii) joint spatial filtering and MCLP based dereverberation in RTF-MCLP is slightly better compared to the cascaded approach of MCLP dereverberation followed by spatial filtering in C-RTF-MCLP, (iii) STOI value is more than $0.9$ for all the acoustic conditions in all three MCLP methods indicating good intelligibility of enhanced speech, and (iv) SDB performance is poor compared to the MCLP based methods. Although all techniques deteriorate with increasing $RT60$, the advantage of RTF-MCLP method is higher for higher $RT60$ than MCLP (Fig. \ref{fig:performanceAcousticParameters}(a-c)). We see that the good performance of RTF-MCLP is over an RT60 range upto $0.7~s$. 
The performance as a function of source distance, shown in \ref{fig:performanceAcousticParameters}(d-f), decreases approximately linearly with increasing distance. Increasing the source distance decreases the SRR (signal to reverberation ratio), which can be akin to increasing $RT60$. In terms of STOI, a source distance of $2.5~m$ has a similar effect as $RT60=1.2~s$. The angular performance plots in \ref{fig:performanceAcousticParameters}(g-i) show that the improvement is approximately uniform for all angular positions of the source, because of the omni-directional microphones assumed.
\subsubsection{Directional speech interference}
\par
\begin{figure}[h]
	\centering
	\includegraphics[width=3.3in,height=3.5in,trim={10pt 0 0 0}]{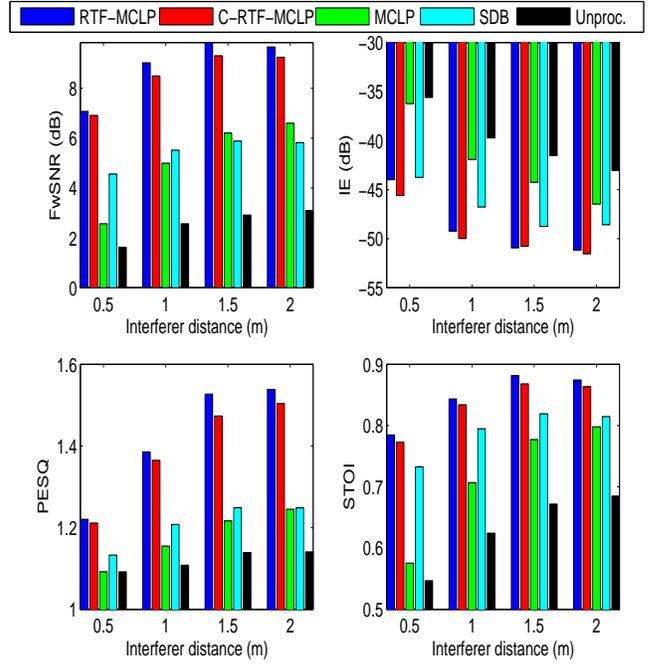}
	\caption{Dereverberation performance for interfering speech source at $45^o$.}
	\label{fig:performanceExternalNoises}
\end{figure}

 Next, we study the performance of the RTF-MCLP method in the presence of an interfering speech source, with $0~dB$ signal to interference ratio (SIR), at $45^o$ angular position. For this experiment we assume the RTF vector of the desired speech source to be known a-priori. It is computed using the discrete Fourier transform of the initial $8~ms$ after the direct component of the desired source RIR. Fig. \ref{fig:performanceExternalNoises} shows the performance measures FwSNR, residual energy of the interferer (IE) in dB, PESQ and STOI, as a function of the distance of the interferer from the center of the array. The quantum of residual inference at the output is determined by passing the known interference signal through the MCLP and MVDR filters estimated using the mixture microphone signal. At a $0.5~m$ distance, the interferer is near to the array than the source and also to the reference microphone at $0^o$ (the source is at $1~m$ distance and $90^o$ angular position). In Fig. \ref{fig:performanceExternalNoises}, we can see that (i) MCLP performance gets severely degraded in the presence of interference, (ii) the super directive beamformer performance is better than MCLP for nearer source positions, i.e., when the direct component is stronger than the reverb components, (iii) cascaded MVDR beamformer at the output of the MCLP filter (C-RTF-MCLP) improves the performance by $3-4~dB$ compared to the beamformer or the MCLP filter alone, (iv) the new RTF-MCLP scheme further improves by about $1~dB$. The PESQ performance is poor for all techniques, but intelligibility is good in terms of STOI, close to $0.9$. The interferer suppression is found to be slightly better in the C-RTF-MCLP approach.
\subsection{Dynamic source dereverberation}
We consider the experimental scenario shown in Fig. \ref{fig:simulationRoomSetup}(b). Both the adaptation parameters $\alpha_1,\alpha_2$ are fixed at $0.1$, and $\epsilon=10^{-6}$. We evaluate the signal estimation performance using segment-wise FwSNR measure. The restoration of short time spectrum is quantified by the segment-wise LLR measure. The segment measures shown are computed every $10~ms$ with $25~ms$ windows and smoothed using triangular window with $1~s$ ($100$ segments) temporal context. We compare results of the online RTF-MCLP approach to the Kalman filter method of \cite{braun2016online}. Time-varying MCLP in the RTF-MCLP method is different from \cite{braun2016online} in the following assumptions: in \cite{braun2016online}, (i) the desired signal spatial covariance matrix is full rank $ \bld{d}[n,k] \sim \mathcal{N}_c \inBrackets{ \bld{0}, \bs{\Phi}_d[n,k]} $, and (ii) the distribution of $\bld{i}_{m,n}[k]$ is an isotropic Gaussian, identically distributed for each $m$, $\bld{g}_{m,n}[k] \sim \mathcal{N}_c \inBrackets{ \bld{0}, \lambda \bld{I}} $. Because of assumption-(i), the Kalman filter approach requires computation of a matrix inverse for each STFT index $n,k$, hence computationally more expensive. Average per frame computation time on a laptop with Intel (R) Core(TM) i5-4210U CPU @ 1.7 GHz processor is found to be $1.07~sec$ and $0.18~sec$ for the Kalman filter approach and the RTF-MCLP respectively providing a ten-fold reduction.
We examine three experimental scenarios in the following, single source with position change, two simultaneous sources and a moving source. Two BBC news clips of duration $\approx 18~s$ are used as test signals.  
\subsubsection{Single source with position switch}
\begin{figure}[h]
	\centering
	\includegraphics[width=3in,height=2.3in,trim={25 0 20 0}]{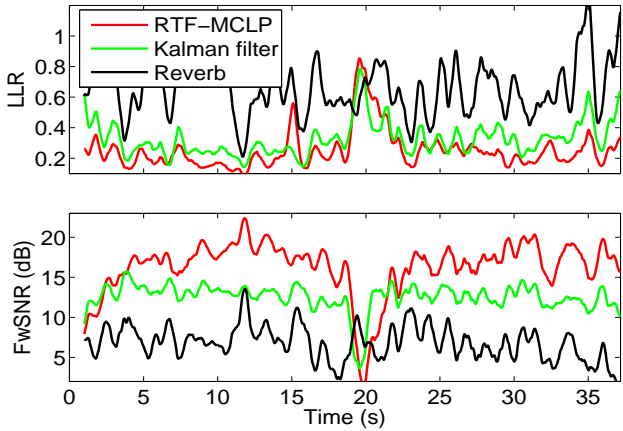}	
	\vspace{-5pt}
	\caption{LLR and FwSNR measures for the source position change scenario.}\label{fig:spc}
\end{figure}

In this experiment, we consider the source to switch from position-A to position-B in Fig. \ref{fig:simulationRoomSetup} at $18.6~s$. This example is to test the adaptation of the online RTF-MCLP to the switched source position. The objective performance measures are shown in Fig. \ref{fig:spc}. We see that, the performance of RTF-MCLP is better in-terms of both LLR and FwSNR measures than the Kalman filter approach, during the steady-state portions. But, the performance does degrade at the source change point (transient response). It is found to be slightly poorer than the Kalman filter approach, probably due to the choice of parameters $\{\alpha_1,\alpha_2, \epsilon\}$, which result in a trade-off between the transient and steady-state performances.
\subsubsection{Stationary, two simultaneous sources}
\begin{figure}[h]
	\centering
	\includegraphics[width=3in,height=3.5in,trim={40 0 20 0}]{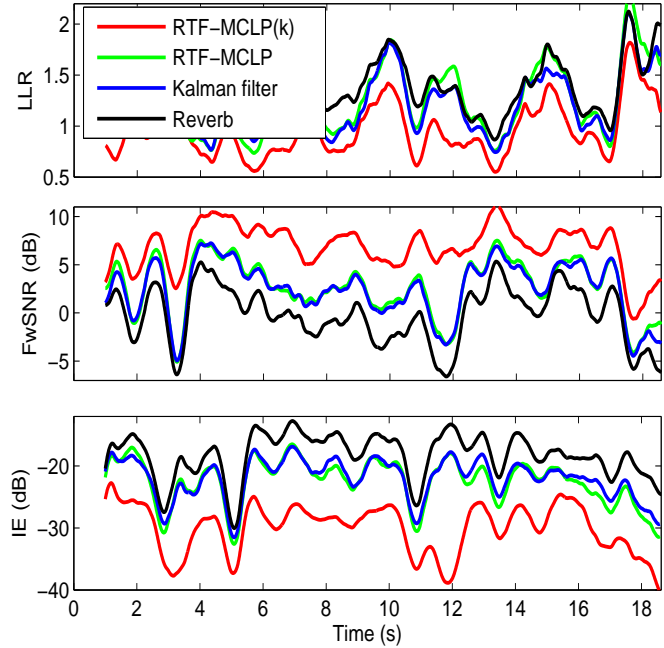}			
	\vspace{-5pt}
	\caption{LLR, FwSNR, and interference energy measures for two simultaneous sources case; RTF-MCLP(k) shows results for the case with RTF of the desired source known a-priori. }\label{fig:tss}
\end{figure}

The performance of estimating a desired source in a two source scenario is shown in Fig. \ref{fig:tss}. Two different BBC news clips are played simultaneously with equal strength ($0~dB$ signal to interference ratio) from positions A and B, shown in Fig. \ref{fig:simulationRoomSetup}(b). The source at position-A is considered as the desired source. RTF estimated from the first $8~ms$ (after direct path) of the RIR is specified for the RTF-MCLP(k) method (i.e., RTF of desired source is known a-priori). The objective performance measures in Fig. \ref{fig:tss} show that, the MCLP based methods are effective even in multi source scenario. The performance of RTF-MCLP is similar to the Kalman filter approach, and shows improvement compared to reverberant speech. The prior fixing of RTF in RTF-MCLP(k) improves the desired signal estimation by $\approx~5~dB$, which can be seen distinctly in the suppression of interference signal energy, FwSNR, and also LLR improvement.
\subsubsection{Moving source}
\begin{figure}[h]
	\centering
	\includegraphics[width=3in,height=2.3in,trim={25 0 20 0}]{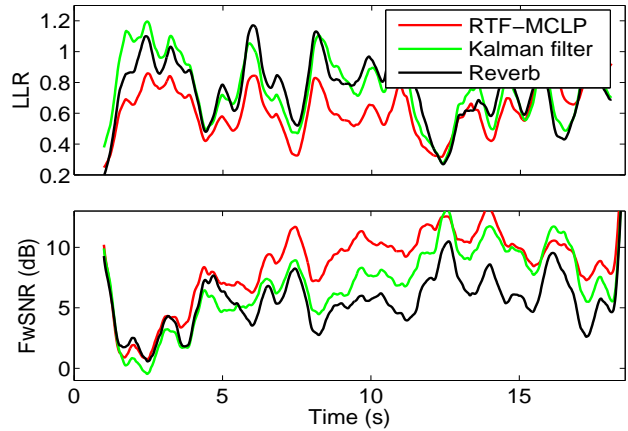}	
	\vspace{-5pt}
	\caption{LLR, and FwSNR measures for moving source experiment.}\label{fig:online}
	\vspace{-5pt}
\end{figure}

We consider a single source moving from position-D to position-C in Fig. \ref{fig:simulationRoomSetup}(b) along a linear path ($2.5~m$ distance) with a fixed speed in $18~s$. The moving source recording is simulated by convolving the source signal with the RIR changing every $5~ms$ (for the chosen positions and duration, this corresponds to $\approx 0.7~mm$ separation between successive positions). For dereverberation, the constants $\alpha_1,\alpha_2$ are set to $0.01$, instead of $0.1$ as in earlier fixed source position case. The results in Fig. \ref{fig:online} show that time-varying MCLP is effective for moving source case as well, and the RTF-MCLP method is better than the Kalman filter approach in terms of both FwSNR and LLR measures. As the distance of source to reference microphone is decreasing with time, we can see that FwSNR improves with time, in both the methods. However, the performance improvement over Kalman filter is less for the moving source case compared to the stationary source cases discussed in the previous section.
\section{Conclusions}
\label{sec:conclusions}
 Joint spatial filtering and multi channel linear prediction in the STFT domain is developed for dereverberation of static or dynamic speech sources in a single source and multiple source applications. For a static source case, batch estimation is considered assuming stochastic time-independence of speech STFT coefficients. The joint formulation of spatial filtering and MCLP is found to provide effective dereverberation, as well as interference suppression with minimum prior knowledge of the source RTF.
The extension of the joint formulation to the moving source case is found to be equally effective because of online spatial filtering and a linear dynamic system model for the MCLP. The interferer suppression is good even at $0~dB$ SIR for a directional separation of $45^o$.


\bibliographystyle{IEEEtran}
\bibliography{references}

\end{document}